\documentclass[pra,aps,superscriptaddress,nofootinbib,twocolumn]{revtex4-1}

\usepackage{mathrsfs}
\usepackage{amsfonts}
\usepackage{amssymb}
\usepackage{amsmath}
\usepackage{graphicx}
\usepackage[usenames,dvipsnames]{color}
\usepackage[colorlinks=true,citecolor=blue,linkcolor=magenta]{hyperref}
\usepackage{ulem}
\usepackage{lmodern}
\usepackage{amsthm}

\newcommand{\figpath}{.}

\newcommand{\Tr}{\mathrm{Tr}}
\newcommand{\norm}[1]{\Vert #1 \Vert}
\newcommand{\abs}[1]{\vert #1 \vert}

\newcommand{\ket}[1]{\vert{ #1 }\rangle}
\newcommand{\bra}[1]{\langle{ #1 }\vert}
\newcommand{\ketbra}[2]{\vert #1 \rangle \langle #2 \vert}
\newcommand{\braket}[2]{\langle #1 \vert #2 \rangle}
\newcommand{\mean}[1]{\langle #1 \rangle}

\newcommand{\LL}{\mathcal{L}}
\newcommand{\PP}{\mathcal{P}}
\newcommand{\KK}{\mathcal{K}}
\newcommand{\kk}{\boldsymbol{\kappa}}
\newcommand{\BB}{\mathcal{B}}
\newcommand{\OO}{\mathcal{O}}
\newcommand{\VV}{\mathcal{V}}

\newtheorem{theorem}{Theorem}

\begin{document}

\title{Hardware-efficient quantum algorithm for the simulation of open-system dynamics and thermalisation}

\author{Hong-Yi Su}
\affiliation{Graduate School of China Academy of Engineering Physics, Beijing 100193, China}

\author{Ying Li}
\email{yli@gscaep.ac.cn}
\affiliation{Graduate School of China Academy of Engineering Physics, Beijing 100193, China}

\begin{abstract}
The quantum open-system simulation is an important category of quantum simulation. By simulating the thermalisation process at the zero temperature, we can solve the ground-state problem of quantum systems. To realise the open-system evolution on the quantum computer, we need to encode the environment using qubits. However, usually the environment is much larger than the system, i.e.~numerous qubits are required if the environment is directly encoded. In this paper, we propose a way to simulate open-system dynamics by reproducing reservoir correlation functions using a minimised Hilbert space. In this way, we only need a small number of qubits to represent the environment. To simulate the $n$-th-order expansion of the time-convolutionless master equation by reproducing up to $n$-time correlation functions, the number of qubits representing the environment is $\sim \lfloor \frac{n}{2} \rfloor \log_2(N_\omega N_\beta)$. Here, $N_\omega$ is the number of frequencies in the discretised environment spectrum, and $N_\beta$ is the number of terms in the system-environment interaction. By reproducing two-time correlation functions, i.e.~taking $n = 2$, we can simulate the Markovian quantum master equation. In our algorithm, the environment on the quantum computer could be even smaller than the system.
\end{abstract}

\maketitle

\section{Introduction}

The idea of quantum computation is motivated by quantum simulation. According to R. Feynman, ``the physical world is quantum mechanical, and therefore the proper problem is the simulation of quantum physics''~\cite{Feynman1982}. The physical world is not only {\it quantum} but also {\it open}. Many vital phenomena are attributed to the open-system dynamics, e.g.~thermalisation~\cite{BreuerPetruccione, Vega2017}. Systems are influenced by their environments through external interactions. Therefore, by simulating the composite system, including the system and the environment, we can study an open system on a quantum computer~\cite{Lloyd1996, Terhal2000, Wang2011}. However, the simulation of the environment is usually inefficient when the environment is big compared to the system. It is also a waste of resources. In many circumstances, we are only interested in the system, not the environment. The simulation of the environment using the most of computational resources may not give us any new knowledge, for instance, when the environment is modeled as exactly solvable boson bath or spin bath. The dynamics of the system is determined by reservoir correlation functions. For example, in the thermalisation, transition rates between eigenstates are determined by two-time reservoir correlation functions~\cite{BreuerPetruccione}. Therefore, reproducing reservoir correlation functions is sufficient, and the full simulation of the environment is unnecessary.

An application of quantum computation is to compute the ground-state energy, which is an important problem in material science and chemistry~\cite{Abrams1999, AspuruGuzik2005, Wecker2014, Bauer2016}. Given an initial state with a finite probability in the ground state, we can use the quantum phase estimation algorithm to obtain the ground-state energy~\cite{Abrams1999, AspuruGuzik2005}. However, we do not have a universal algorithm that can prepare such an initial state~\cite{Verstraete2009, Farhi2001}. Solving the ground-state problem for a general Hamiltonian is likely to be intractable even in quantum computation~\cite{Kitaev2002, Aharonov2002}. A related problem is preparing or sampling thermal states of a quantum system~\cite{Poulin2009, Bilgin2010, Temme2011, Riera2012, Yung2012, Motta2019}, and the ground state is the thermal state at the zero temperature. If we only focus on systems in the real world, most of them reach the thermal state as a result of the open-system dynamics. Therefore, for such real-world systems, simulating the open-system dynamics is an efficient way to prepare thermal states, including the ground state. Although we have quantum algorithms that can implement semi-group dynamics (unitary or non-unitary)~\cite{Lloyd1996, Berry2007, Wiebe2010, Berry2015, Campbell2019, Bacon2001, Kliesch2011, Sweke2015, Candia2015, Sweke2016, Childs2017, Chenu2017}, they cannot be directly used for the thermalisation by simulating the corresponding Lindblad equation. Working out the Lindblad equation of the thermalisation requires the spectrum of the system~\cite{BreuerPetruccione}, which is the information that we want to obtain in the computation. To computation the thermal state, we have to assume that the Lindblad equation is unknown. Therefore, we need an environment to simulate the thermalisation. In this paper, we propose a hardware-efficient quantum algorithm for the simulation of Markovian and non-Markovian open-system dynamics, which can be used for solving the thermalisation and ground-state problems.

Qubits are valuable resources in the present and future. It is similar to the classical computational resources we use today but more severe. Fault-tolerant quantum computation based on the quantum error correction is the way to implement large-scale quantum computations, in which encoding one logical qubit may need thousands of physical qubits~\cite{Fowler2012, OGorman2017}. Therefore, reducing the number of logical qubits is essential. Variational quantum algorithms for solving the ground-state problem or simulating the real and imaginary time evolutions have been developed recently~\cite{Peruzzo2014, Wecker2015, Li2017, McArdle2018}, which can avoid the enormous qubit cost and are suitable for the near-term quantum computation. In this paper, our algorithm is in the category of conventional quantum algorithms demanding fault tolerance but does not rely on a good variational ansatz. We reduce the qubit cost by using a small environment to simulate the open-system dynamics induced by a big environment. We achieve it by reproducing reservoir correlation functions of the big environment in the small environment.

Open-system dynamics is determined by reservoir correlation functions. According to the expansion of the time-convolutionless (TCL) master equation, the simulation of open-system dynamics is more accurate if higher-order correlation functions are reproduced. By reproducing two-time correlation functions, we can simulate the Redfield equation and, therefore, the Markovian quantum master equation when the Markov approximation is justified. From the Markovian quantum master equation, we can simulate the thermalisation.

Our algorithm is beyond two-time correlation functions. Any $n$-time correlation functions can be reproduced, therefore we can simulate TCL master equation up to any $n$-th-order expansion. Reservoir correlation functions can be reproduced using tensor network, in which way the dimension of the environment increases exponentially with the number of terms in the system-environment coupling~\cite{Luchnikov2019}. Our algorithm uses a different approach. By minimising the Hilbert space dimension for reproducing given correlation functions, the number of qubits required for representing the environment is $\sim \frac{n}{2}\log_2(N_\omega N_\beta)$, where $N_\omega$ is the number of frequencies in the discretised environment spectrum, and $N_\beta$ is the number of terms in the coupling. Reservoir correlation functions can be exactly reproduced up to the spectrum discretisation, which usually converges polynomially with $N_\omega$.

\begin{figure}[tbp]
\centering
\includegraphics[width=1\linewidth]{\figpath 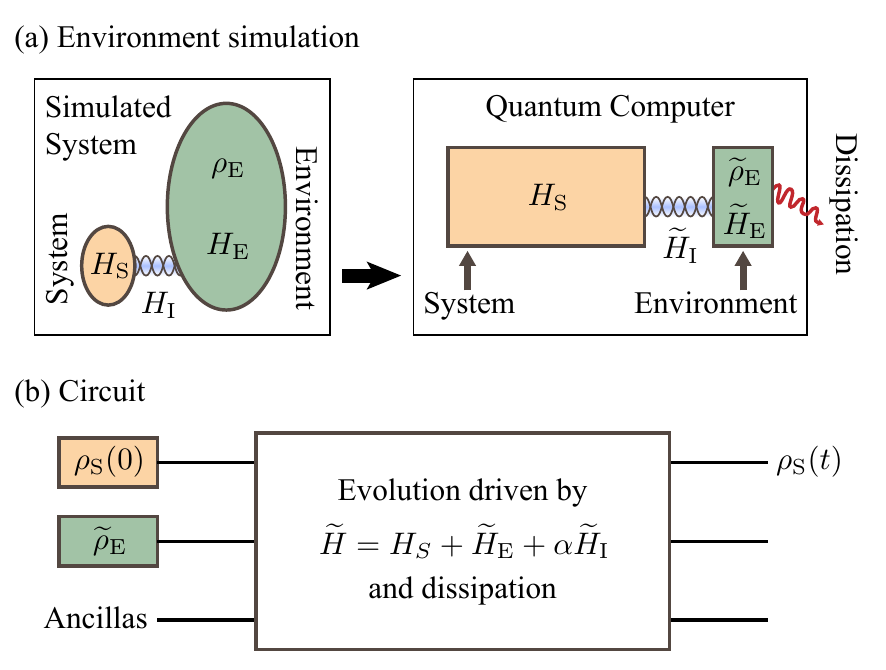}
\caption{
(a) The simulated dynamics of the system is determined by the Hamiltonian $H = H_{\rm S} + H_{\rm E} + \alpha H_{\rm I}$ and the state of the environment $\rho_{\rm E}$. On the quantum computer, instead of directly simulating the environment, we use a Hilbert space with a much lower dimension to represent the environment and simulate the dynamics driven by the Hamiltonian $\widetilde{H} = H_{\rm S} + \widetilde{H}_{\rm E} + \alpha \widetilde{H}_{\rm I}$ and the environment state $\widetilde{\rho}_{\rm E}$. Because the environment on the quantum computer has a finite size, we may need to introduce dissipation in order to relax the environment.
(b) The evolution driven by $\widetilde{H}$ and the dissipation is realised using a quantum circuit on the quantum computer. To obtain the final state of the system $\rho_{\rm S}(t)$, qubits representing the system are prepared in the initial state of the system $\rho_{\rm S}(0)$, qubits representing the environment are prepared in the initial state of the environment $\widetilde{\rho}_{\rm E}$, and then the evolution is implemented using computation operations, i.e.~state preparation, quantum gates and measurement operation. Ancillary qubits may be needed in the simulation, e.g.~for implementing the dissipation.
}
\label{fig:algorithm}
\end{figure}

The theory of open-system dynamics is introduced in Sec.~\ref{sec:open-system}. To simulate the open system dynamics given by the Hamiltonian $H$ and the environment state $\rho_{\rm E}$, instead, we implement the dynamics of the Hamiltonian $\widetilde{H}$ and the environment state $\widetilde{\rho}_{\rm E}$ on the quantum computer, as shown in Fig.~\ref{fig:algorithm}. The overview of the algorithm is given in Sec.~\ref{sec:overview}, and details of the algorithm are discussed in Sec.~\ref{sec:Markovian}, \ref{sec:space} and \ref{sec:algorithm}. In Sec.~\ref{sec:reinitialisation}, we discuss how to reinitialise the environment in the simulation. In Sec.~\ref{sec:cost}, the circuit implementation, qubit cost and gate-number cost are discussed. In Sec.~\ref{sec:thermalisation}, we give an illustrative example, and we numerically implement our algorithm on a classical computer to simulate the thermalisation of a qubit.

Later we will show how to choose $\widetilde{H}$ and $\widetilde{\rho}_{\rm E}$ such that reservoir correlation functions of $H$ and $\rho_{\rm E}$ can be reproduced on the quantum computer. Because we want to minimise the number of qubits representing the environment, the state of the environment may significantly change with time. Therefore, we may need to re-initialise the environment state during the simulation. We also show how to implement the re-initialise without significantly modifying correlation functions.

\section{Dynamics of open quantum systems weakly coupled to the environment}
\label{sec:open-system}

Given the Hamiltonian of the system and environment $H = H_{\rm S} + H_{\rm E} + \alpha H_{\rm I}$, where $H_{\rm S}$, $H_{\rm E}$ and $H_{\rm I}$ respectively denote Hamiltonians of the system, environment and interaction [see Fig.~\ref{fig:algorithm}(a)], the evolution equation in the interaction picture is
\begin{eqnarray}
\frac{\partial}{\partial t} \rho(t) = -i \alpha[H_{\rm I}(t),\rho(t)] \equiv \alpha\LL(t)\rho(t).
\end{eqnarray}
Here, $\rho$ is the state of the system and environment, $\alpha$ is a dimensionless coupling constant, and we have taken $\hbar = 1$. Derived from this evolution equation, TCL equation~\cite{BreuerPetruccione} is
\begin{eqnarray}
\frac{\partial}{\partial t} \PP\rho(t) = \KK(t)\PP\rho(t)
\end{eqnarray}
for any initial state in the form $\rho(0) = \rho_{\rm S}(0)\otimes\rho_{\rm E}$, where $\PP$ is a superoperator projection defined by $\PP\rho \equiv \Tr_{\rm E}(\rho)\otimes\rho_{\rm E}$. We focus on the case that $\rho_{\rm E}$ is a stationary state of the environment, i.e.~$[H_{\rm E}, \rho_{\rm E}] = 0$.

TCL master equation is the evolution equation of the system state $\rho_{\rm S} = \Tr_{\rm E}(\rho)$, because $\PP\rho(t) = \rho_{\rm S}(t)\otimes\rho_{\rm E}$, in which only the system state $\rho_{\rm S}(t)$ evolves with time, and the environment state $\rho_{\rm E}$ is constant.

When the coupling between the system and environment is weak, the expansion of TCL generator $\KK(t)$ in powers of the coupling constant $\alpha$ provides a series of approximate evolution equations. The expansion reads $\KK(t) = \sum_{n=1}^\infty \alpha^n \KK_n(t)$, where $\KK_n(t)$ does not depend on $\alpha$. For example, up to the fourth order, we have $\KK_1(t) = \PP\LL(t)\PP$, $\KK_2(t) = \int_0^{t} dt_1 \PP\LL(t)\LL(t_1)\PP$, $\KK_3(t) = \int_0^{t} dt_1 \int_0^{t_1} dt_2 \PP\LL(t)\LL(t_1)\LL(t_2)\PP$ and
\begin{eqnarray}
\KK_4(t) &=& \int_0^{t} dt_1 \int_0^{t_1} dt_2 \int_0^{t_2} dt_3 \notag \\
&& \big[ \kk_4(t,t_1,t_2,t_3) - \kk_2(t,t_1)\kk_2(t_2,t_3) \notag \\
&& - \kk_2(t,t_2)\kk_2(t_1,t_3) - \kk_2(t,t_3)\kk_2(t_1,t_2) \big],
\label{eq:K4}
\end{eqnarray}
where
\begin{eqnarray}
\kk_n(t,t_1,\ldots,t_{n-1}) = \PP\LL(t)\LL(t_1)\cdots\LL(t_{n-1})\PP.
\end{eqnarray}
Without loss of generality, we have assumed $\PP\LL(t)\PP = 0$ for simplification. In general, the $n$-th-order TCL generator $\KK_n(t)$ is determined by superoperators $\kk_m$ with $m \leq n$.

TCL equation is the exact evolution equation of the system, therefore, describes the non-Markovian dynamics of the system. By neglecting high-order terms and taking the approximation $\KK_2(t) \simeq \KK_2(\infty)$ under the assumption that the correlation time is short, we can get the Markovian quantum master equation $\frac{\partial}{\partial t} \PP\rho(t) = \alpha^2\KK_2(\infty)\PP\rho(t)$.

\subsection*{Correlation functions of the environment}

Open-system dynamics is determined by reservoir correlation functions. In general, the $n$-th-order TCL generator $\KK_n(t)$ is determined by up to the $n$-time correlations~\cite{BreuerPetruccione}. The interaction can always be expressed in the form $H_{\rm I} = \sum_\beta A_\beta \otimes B_\beta$, where $A_\beta$ acts on the system, $B_\beta$ acts on the environment, and they are both Hermitian. Expanding $\kk_n$ using the expression of $H_{\rm I}$, we have
\begin{eqnarray}
\kk_n(t,\ldots,t_{n-1}) \rho &&= \sum_{\nu,\ldots,\nu_{n-1}} \sum_{\beta,\ldots,\beta_{n-1}}
i^n (-1)^{\nu+\cdots+\nu_{n-1}} \notag \\
&\times & \Tr \big[ \BB_{\beta}(t,\nu) \cdots \BB_{\beta_{n-1}}(t_{n-1},\nu_{n-1}) \rho_{\rm E} \big] \notag \\
&\times & \mathcal{A}_{\beta}(t,\nu) \cdots \mathcal{A}_{\beta_{n-1}}(t_{n-1},\nu_{n-1}) \PP \rho.
\label{eq:corr_fun}
\end{eqnarray}
Here, $\nu,\ldots,\nu_{n-1} = 0,1$ are binary numbers indicating on which side the Hamiltonian acts, we define superoperators $\mathcal{C}_{\beta}(\iota, \nu) \rho \equiv [C_{\beta}(\iota)]^{\nu} \rho [C_{\beta}(\iota)]^{1-\nu}$, $C = A,B$, and $\iota = t, \omega$ ($\iota = \omega$ will be used later). Therefore, the superoperator $\kk_n$ is determined by system operators $A_\beta$ and $n$-time correlation functions of the environment. We note that reservoir correlation functions are time-ordered, i.e.~$t \geq \cdots \geq t_{n-1}$.

For the environment with a discretised spectrum, the environment Hamiltonian can be decomposed according to the spectrum as $H_{\rm E} = \sum_\varepsilon \varepsilon \Pi(\varepsilon)$, where $\Pi(\varepsilon)$ is the projection onto the eigenspace of the eigenenergy $\varepsilon$. We define operators $B_\beta(\omega) \equiv \sum_{\varepsilon'-\varepsilon = \omega} \Pi(\varepsilon) B_\beta \Pi(\varepsilon')$, then $B_{\beta}(t) = \sum_\omega e^{-i\omega t} B_\beta(\omega)$. We can find that $B_\beta^\dag(\omega) = B_\beta(-\omega)$, because $B_\beta$ is Hermitian.

Without loss of generality, we assume $\Tr (B_\beta \rho_{\rm E}) = 0$, i.e.~$\PP\LL(t)\PP = 0$. We note that $\rho_{\rm E}$ is a stationary state. If $\Tr (B_\beta \rho_{\rm E})$ is not zero, we can replace $B_\beta$ with $B_\beta - \Tr (B_\beta \rho_{\rm E})\openone$ and $H_{\rm S}$ with $H_{\rm S} + \sum_\beta \Tr (B_\beta \rho_{\rm E}) A_\beta \otimes \openone$, so that the total Hamiltonian is not changed but the assumption is satisfied.

Two distinct environments result in the same dynamics of the system if their correlation functions are the same and they are coupled to the system by the same set of operators $A_\beta$ [see Fig.~\ref{fig:algorithm}(a)].

\begin{theorem}
Let $\widetilde{H} = H_{\rm S} + \widetilde{H}_{\rm E} + \alpha \widetilde{H}_{\rm I}$ be the Hamiltonian of the system and a different environment, and $\widetilde{H}_I = \sum_\beta A_\beta \otimes \widetilde{B}_\beta$. The sufficient condition for the same dynamics of the system up to the $n$-th-order, i.e.~$\sum_{m=1}^n \alpha^m \KK_m(t) = \PP \sum_{m=1}^n \alpha^m \widetilde{\KK}_m(t)$, is that
\begin{eqnarray}
&& \Tr \big[ \widetilde{\BB}_{\beta}(t,\nu) \cdots \widetilde{\BB}_{\beta_{m-1}}(t_{m-1},\nu_{m-1}) \widetilde{\rho}_{\rm E} \big] \notag \\
&=& \Tr \big[ \BB_{\beta}(t,\nu) \cdots \BB_{\beta_{m-1}}(t_{m-1},\nu_{m-1}) \rho_{\rm E} \big]
\label{eq:BBrho}
\end{eqnarray}
holds for all $m\leq n$.
\label{the:simulation}
\end{theorem}

\section{Overview of the algorithm}
\label{sec:overview}

According to Theorem~\ref{the:simulation}, in order to simulate the open-system dynamics driven by $H$ up to the $n$-th order, we can implement the evolution driven by $\widetilde{H}$ on the quantum computer. The algorithm has two stages. At the first stage, we compute correlation functions of the environment determined by $H_{\rm E}$ and $\{B_\beta\}$ and design the environment on the quantum computer, i.e.~choose $\widetilde{H}_{\rm E}$, $\{\widetilde{B}_\beta\}$ and the dissipation of the environment to reproduce the same correlation functions. The purpose of our algorithm is to simulate the dynamics of an open quantum system and study the system rather than the environment. We assume that correlation functions of the environment are computable in classical computation. If correlation functions of the environment are classically intractable, we may need the quantum computer to study the dynamics of the environment, which is beyond the scope of this work. On the quantum computer, we want to minimise the size of the environment, therefore dissipation of the environment may be required in order to relax the environment and suppress the finite-size effect. We will give the protocol for designing the environment on the quantum computer later. At the second stage, we use the quantum computer to realise the time evolution driven by $\widetilde{H}$ and the dissipation~[see Fig.~\ref{fig:algorithm}(b)]. Given the corresponding Lindblad equation in the explicit form, the evolution can be realised on the quantum computer using a quantum circuit~\cite{Bacon2001, Kliesch2011, Sweke2015, Candia2015, Sweke2016, Childs2017, Chenu2017}.

\section{Simulation of the second-order equation and Markovian master equation}
\label{sec:Markovian}

In this section, we consider the quantum simulation of the master equation with the second-order approximation. If $\PP\LL(t)\PP = 0$ and higher-order contributions are neglected, the evolution equation of the system reads $\frac{\partial}{\partial t} \PP\rho(t) = \alpha^2 \KK_2(t)\PP\rho(t)$, which can also be expressed in the from
\begin{eqnarray}
\frac{d\rho_{\rm S}}{dt} = -\alpha^2 \int_0^t ds \Tr_{\rm E} [H_{\rm I}(t), [H_{\rm I}(t-s), \rho_{\rm S}\otimes \rho_{\rm E}]].~~~~~
\label{eq:2ndQME}
\end{eqnarray}
This equation is determined by two-time correlation functions
\begin{eqnarray}
\mean{ B_{\beta}(t) B_{\beta'}(t-s) } \equiv \Tr \big[ B_{\beta}(t) B_{\beta'}(t-s) \rho_{\rm E} \big].
\end{eqnarray}
In order to simulate the time evolution driven by Eq.~(\ref{eq:2ndQME}), we reproduce such correlation functions on the quantum computer.

When the time scale over which reservoir correlation functions decay is negligible compared to the time scale over which the system evolves significantly, the Markov approximation is justified. Then $\KK_2(t) \simeq \KK_2(\infty)$ and the upper limit $t$ of the integral in Eq.~(\ref{eq:2ndQME}) can be replaced by $\infty$. Our algorithm can simulate the open system dynamics with a finite correlation time, i.e.~the dynamics is non-Markovian, but we focus on the case that the correlation time is short although may not be negligible.

\subsection{Algorithm for the second-order simulation}

The correlation function of the environment can be expressed in the form
\begin{eqnarray}
\mean{ B_{\beta}(t) B_{\beta'}(t-s) } = \sum_{\omega} e^{-i\omega s} \gamma_{\beta,\beta'}(\omega).
\end{eqnarray}
where $\gamma_{\beta,\beta'}(\omega) = \Tr \big[ B_{\beta}(\omega) B_{\beta'}^\dag(\omega) \rho_{\rm E} \big]$. To choose the interaction operators $\{\widetilde{B}_\beta\}$ on the quantum computer, we diagonalise matrices $\gamma(\omega)$ on a classical computer. Matrices $\gamma(\omega)$ are Hermitian and positive. After the diagonalisation, we obtain $\gamma(\omega) = U(\omega) \Lambda(\omega) U^\dag(\omega)$, where $\Lambda(\omega)$ is the diagonalised matrix, and $U(\omega)$ is unitary. Interaction operators $\{\widetilde{B}_\beta\}$ on the quantum computer depend on coefficients $g_{\beta,l}(\omega) = U_{\beta,l}(\omega) \sqrt{\Lambda_{l,l}(\omega)}$.

On the quantum computer, we use the Hilbert space $\widetilde{\mathcal{H}}_{\rm E} = \widetilde{\mathcal{H}}_{\rm v} \oplus \bigoplus_\omega \widetilde{\mathcal{H}}_\omega$ to represent the environment. Here, $\widetilde{\mathcal{H}}_{\rm v}$ is one-dimensional and contains only one state $\ket{\rm v}$ representing the vacuum, $\widetilde{\mathcal{H}}_\omega$ is $d_\omega$-dimensional and corresponds to the transition frequency $\omega$, and $d_\omega = \mathrm{rank}(\gamma(\omega))$. The orthonormal basis of $\widetilde{\mathcal{H}}_\omega$ is $\{ \ket{\omega, l} \}$, where $l$ corresponds to the $l$-th eigenvalue of $\gamma(\omega)$. The dimension of the environment $\widetilde{\mathcal{H}}_{\rm E}$ is $d_{\rm E} = 1+\sum_\omega d_\omega$, therefore we can use $N_{\rm E} = \lceil \log_2 d_{\rm E} \rceil$ qubits to simulate the environment. We have $d_\omega \leq N_\beta$ and $d_{\rm E} \leq 1 + N_\omega N_\beta$, where $N_\beta \equiv \abs{\{ \beta \}}$ is the number of terms in the interaction Hamiltonian, $N_\omega \equiv \abs{\{ \omega \}} \leq N_\varepsilon^2$ is the number of transition frequencies, and $N_\varepsilon \equiv \abs{\{ \varepsilon \}}$ is the number of eigenenergies in the discretised spectrum of the environment.

To simulate the environment, we take $\widetilde{\rho}_{\rm E} = \ketbra{\rm v}{\rm v}$,
\begin{eqnarray}
\widetilde{H}_{\rm E} &=& \sum_{\omega,l} \omega \sigma_{l}^\dag(\omega) \sigma_{l}(\omega), \\
\widetilde{B}_\beta &=& \sum_{\omega,l} g_{\beta,l}(\omega) \sigma_{l}(\omega) + {\rm h.c.}, \label{eq:B2nd}
\end{eqnarray}
where $\sigma_{l}(\omega) \equiv \ketbra{{\rm v}}{\omega, l}$. Then, correlations functions can be reproduced on the quantum computer. We note that $\Tr \big[ \widetilde{B}_{\beta}(\omega) \widetilde{B}_{\beta'}^\dag(\omega) \widetilde{\rho}_{\rm E} \big] = [U(\omega) \sqrt{\Lambda(\omega)} \sqrt{\Lambda(\omega)} U^\dag(\omega)]_{\beta,\beta'} = \gamma_{\beta,\beta'}(\omega)$, where
\begin{eqnarray}
\widetilde{B}_{\beta}(\omega) = \sum_{l} g_{\beta,l}(\omega) \sigma_{l}(\omega) + g_{\beta,l}^*(-\omega) \sigma_{l}^\dag(-\omega).
\end{eqnarray}
Therefore, $\mean{ \widetilde{B}_{\beta}(t) \widetilde{B}_{\beta'}(t-s) } = \mean{ B_{\beta}(t) B_{\beta'}(t-s) }$, where $\mean{ \widetilde{B}_{\beta}(t) \widetilde{B}_{\beta'}(t-s) } = \Tr \big[ \widetilde{B}_{\beta}(t) \widetilde{B}_{\beta'}(t-s) \widetilde{\rho}_{\rm E} \big]$.

\subsection{Discussion}

\begin{figure}[tbp]
\centering
\includegraphics[width=1\linewidth]{\figpath 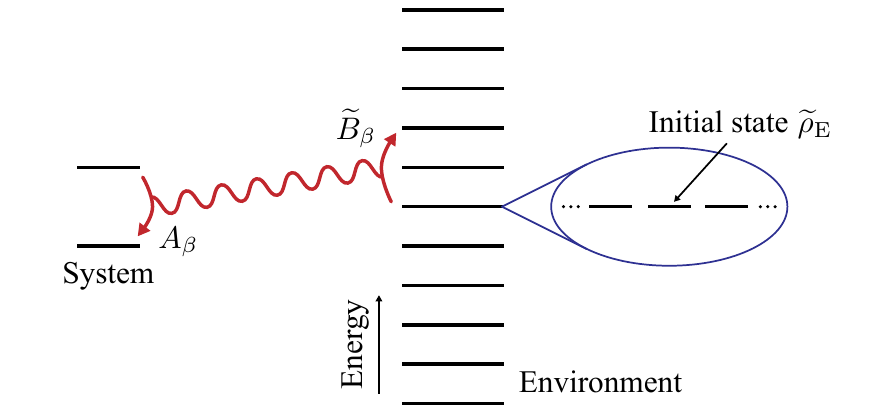}
\caption{
Level scheme of the environment simulation. Each energy level of the environment represents a frequency with respect to the initial state of the environment. These levels are degenerate. The initial state is encoded as a pure state on the level with the frequency $0$. The system can release energy into the environment or absorb energy from the environment via transitions between these energy levels. Transitions are caused by the interaction $\widetilde{H}_I = \sum_\beta A_\beta \otimes \widetilde{B}_\beta$.
}
\label{fig:level}
\end{figure}

In the algorithm, the initial state of the environment on the quantum computer is always the pure state $\ket{\rm v}$, and the pure state is not the ground state, because the frequency $\omega$ can take negative values (see Fig.~\ref{fig:level}). The system can release energy into the environment via a transition from the state $\ket{\rm v}$ to the state $\ket{{\omega, l}}$ with a positive $\omega$. Similarly, the system can absorb energy from the environment via a transition from the state $\ket{\rm v}$ to the state $\ket{{\omega, l}}$ with a negative $\omega$. For a thermal bath with the temperature $T$, the $\gamma$ matrix satisfies $\gamma_{\beta,\beta'}(-\omega) = \exp(-\hbar\omega/k_{\rm B}T) \gamma_{\beta',\beta}(\omega)$~\cite{BreuerPetruccione}. If the temperature is $0$, $\gamma(\omega) = 0$, i.e.~$g_{\beta,l}(\omega) = 0$, for all negative $\omega$. Then, states $\ket{{\omega, l}}$ with a negative $\omega$ are decoupled from the system. In this case, the state $\ket{\rm v}$ is the effective ground state. If the temperature is finite, the system is not only coupled to positive-$\omega$ states but also negative-$\omega$ states. In this way, we can simulate a finite temperature environment using a pure state as the initial state of the environment.

\section{State space relevant to the $n$-th-order expansion}
\label{sec:space}

We can generalise the algorithm in Sec.~\ref{sec:Markovian} to simulate the TCL master equation up to the $n$-th order. Before giving the general algorithm, we first analyse the space of relevant environment states. The dimension of the state space has an upper bound $d_{n,{\rm max}} = [(N_\omega N_\beta)^{\lfloor n/2 \rfloor+1} -1]/[N_\omega N_\beta -1]$ as we explain in the next paragraph. Given the dimension of the state space, we can use a Hilbert space with the same dimension as the environment on the quantum computer to simulate the $n$-th-order TCL master equation.

Now we explain the upper bound of the dimension. We can rewrite the $m$-time correlation function as
\begin{eqnarray}
&& \Tr \big[ \BB_{\beta}(t,\nu) \cdots \BB_{\beta_{m-1}}(t_{m-1},\nu_{m-1}) \rho_{\rm E} \big] \notag \\
&=& \sum_{\omega,\ldots,\omega_{m-1}} e^{-i(\omega t + \cdots + \omega_{m-1} t_{m-1})} \notag \\
&&\times \Tr \big[ \BB_{\beta}(\omega,\nu) \cdots \BB_{\beta_{m-1}}(\omega_{m-1},\nu_{m-1}) \rho_{\rm E} \big].
\label{eq:transformation}
\end{eqnarray}
Let $\ket{\psi}$ be the purification of the state $\rho_{\rm E}$, i.e.~$\ket{\psi}$ is a state on the Hilbert space $\mathcal{H}_{\rm E}\otimes\mathcal{H}_{\rm a}$ satisfying $\Tr_{\rm a} (\ketbra{\psi}{\psi}) = \rho_{\rm E}$. Here, $\mathcal{H}_{\rm E}$ is the Hilbert space of the environment, and $\mathcal{H}_{\rm a}$ is the Hilbert space of an ancillary system with the minimum dimension $\mathrm{rank}(\rho_{\rm E})$. We define $b_\beta(\omega) \equiv B_\beta(\omega) \otimes \openone_{\rm a}$, where $\openone_{\rm a}$ is the identity operator of $\mathcal{H}_{\rm a}$. Expressing superoperators $\BB$ using operators $B$, the last line of Eq.~(\ref{eq:transformation}) can be expressed in the form
$$\Tr[ \text{(some }B\text{ operators)} \rho_{\rm E} \text{(some }B\text{ operators)} ],$$
and we can rewrite it as
$$\Tr[ \text{(some }b\text{ operators)} \ketbra{\psi}{\psi} \text{(some }b\text{ operators)} ].$$
Then using the cyclic property of trace, we rewrite correlation functions in the form
\begin{eqnarray}
&& \Tr \big[ \BB_{\beta}(\omega,\nu) \cdots \BB_{\beta_{m-1}}(\omega_{m-1},\nu_{m-1}) \rho_{\rm E} \big] \notag \\
&=& \bra{\psi} [b_{\beta_{m-1}}(\omega_{m-1})]^{1-\nu_{m-1}} \cdots [b_{\beta}(\omega)]^{1-\nu} \notag \\
&&\times [b_{\beta}(\omega)]^{\nu} \cdots [b_{\beta_{m-1}}(\omega_{m-1})]^{\nu_{m-1}} \ket{\psi}.
\label{eq:rho2psi}
\end{eqnarray}
To simulate the $n$-th-order TCL master equation, we only need to consider correlation functions with $m \leq n$. Therefore, the maximum number of $b$ operators in the above equation is $n$. We introduce states
\begin{eqnarray}
&& \ket{\phi_{\Omega,m}(\beta, \omega, \beta_1, \omega_1, \ldots, \beta_{m-1}, \omega_{m-1})} \notag \\
&\equiv & b_{\beta_{m-1}}(\omega_{m-1}) \cdots b_{\beta_1}(\omega_1) b_{\beta}(\omega) \ket{\psi},
\end{eqnarray}
where $\Omega = \omega + \omega_1 + \cdots + \omega_{m-1}$. Then, correlation functions can always be expressed in the form
\begin{eqnarray}
&& \Tr \big[ \BB_{\beta}(\omega,\nu) \cdots \BB_{\beta_{m-1}}(\omega_{m-1},\nu_{m-1}) \rho_{\rm E} \big] \notag \\
&=& \bra{\phi_{\Omega_{\rm L},m_{\rm L}}(\cdots)} [b_{\beta'}(\omega')]^{\nu'} \ket{\phi_{\Omega_{\rm R},m_{\rm R}}(\cdots)},
\label{eq:rho2phi}
\end{eqnarray}
where two arguments of $\phi$, $\beta'$, $\omega'$ and $\nu'$ on the second line depends on $\beta$, $\omega$ and $\nu$ one the first line. Here, we remark that $b_{\beta}^\dag(\omega) = b_{\beta}(-\omega)$. Because the maximum number of $b$ operators is $n$, all correlations can be expressed in the above form with $m_{\rm L},m_{\rm R}\leq \lfloor n/2 \rfloor$. The number of states $\ket{\phi_{\Omega,m}(\cdots)}$ is $(N_\omega N_\beta)^m$, because there are $m$ operators $b$ acting on $\ket{\psi}$, and each operator $b$ has $N_\omega N_\beta$ options. Therefore, the total number of all relevant states $V_n = \{ \ket{\phi_{\Omega,m}(\cdots)} ~\vert~ m \leq \lfloor n/2 \rfloor \}$ is $1 + N_\omega N_\beta + \cdots + (N_\omega N_\beta)^{\lfloor n/2 \rfloor}$, which is the upper bound of the space dimension.

We can decompose the space of relevant states according to the frequency. Because $\rho_{\rm E}$ is a stationary state, the correlation function in Eq.~(\ref{eq:rho2phi}) is nonzero only if the summation of frequencies is zero, i.e.~$\omega + \omega_1 + \cdots + \omega_{m-1} = 0$. Therefore, two states $\ket{\phi_{\Omega_{\rm R},m_{\rm R}}(\cdots)}$ and $\ket{\phi_{\Omega_{\rm L},m_{\rm L}}(\cdots)}$ are orthogonal if $\Omega_{\rm R}\neq \Omega_{\rm L}$. Then, the space of relevant states can be decomposed as $\mathcal{H}_{\rm E}^{\rm r} = \bigoplus_\Omega \mathcal{H}_\Omega$, where $\mathcal{H}_\Omega$ is the span of states $\{ \ket{\phi_{\Omega,m}(\cdots)} \}$ with the frequency $\Omega$.

\section{General algorithm for simulating the environment}
\label{sec:algorithm}

The algorithm has two stages. At the first stage, we compute correlation functions of the environment and work out how to encode the environment on the quantum computer. At the second stage, we use the quantum computer to realise the time evolution driven by a Hamiltonian worked out at first stage.

\subsection{Classical computation}

To simulate the dynamics of an open quantum system up to the $n$-th-order expansion of TCL equation, we compute correlation functions, $g_{\phi,\phi'} = \braket{\phi}{\phi'}$ and $b_{\phi,\phi'} = \bra{\phi} b \ket{\phi'}$, where $\ket{\phi}, \ket{\phi'} \in V_n$ and $b\in \{b_{\beta}(\omega)\}$. These correlations functions are all in the form of the last line in Eq.~(\ref{eq:rho2phi}).

Using the Gram matrix $g_{\phi,\phi'}$ and Gram-Schmidt orthogonalisation (see Appendix), we can obtain a $d_{\rm E}$-dimensional representation of states $\ket{\phi}$ and operators $b$, where $d_{\rm E} = \mathrm{rank}(g) \leq d_{n,{\rm max}}$ is the dimension of $\mathcal{H}_{\rm E}^{\rm r}$. Each state $\ket{\phi} \in V_n$ maps to a $d_{\rm E}$-dimensional vector $\ket{\widetilde{\phi}}$, and each $b\in \{b_{\beta}(\omega)\}$ maps to a $d_{\rm E}$-dimensional matrix $\widetilde{b}$. These $d_{\rm E}$-dimensional vectors and matrices satisfy $\braket{\widetilde{\phi}}{\widetilde{\phi}'} = g_{\phi,\phi'}$ and $\bra{\widetilde{\phi}} \widetilde{b} \ket{\widetilde{\phi}'} = b_{\phi,\phi'}$. Then,
\begin{eqnarray}
\bra{\widetilde{\psi}} \widetilde{b}_m \cdots \widetilde{b}_2 \widetilde{b}_1 \ket{\widetilde{\psi}} = \bra{\psi} b_m \cdots b_2 b_1 \ket{\psi}
\label{eq:vbbbv}
\end{eqnarray}
holds for all $m$-th order correlation functions if $m\leq 2\lfloor n/2 \rfloor + 1$. Given $\ket{\widetilde{\phi}}$ and $\widetilde{b}$, we can simulate dynamics of the open quantum system on the quantum computer.

As the same as $\mathcal{H}_{\rm E}^{\rm r}$, the space of vectors $\{ \ket{\widetilde{\phi}} \}$ can be decomposed in the form $\widetilde{\mathcal{H}}_{\rm E} = \bigoplus_\Omega \widetilde{\mathcal{H}}_\Omega$, where $\widetilde{\mathcal{H}}_\Omega$ is the span of states $\{ \ket{\widetilde{\phi}_{\Omega,m}(\cdots)} \}$ with the frequency $\Omega$, because $\braket{\widetilde{\phi}_{\Omega_{\rm L},m_{\rm L}}(\cdots)}{\widetilde{\phi}_{\Omega_{\rm R},m_{\rm R}}(\cdots)} = 0$ if $\Omega_{\rm R} \neq \Omega_{\rm L}$. We remark that $\ket{\widetilde{\psi}}$ is in the subspace $\widetilde{\mathcal{H}}_{\Omega =0}$.

\subsection{Quantum computation}

The simulation performed on the quantum computer is as follows. On the quantum computer, we use a $d_{\rm E}$-dimensional Hilbert space $\widetilde{\mathcal{H}}_{\rm E} = \bigoplus_\Omega \widetilde{\mathcal{H}}_\Omega$, i.e.~$N_{\rm E} = \lceil \log_2 d_{\rm E} \rceil$ qubits, to represent the environment, where $d_{\rm E} \leq d_{n,{\rm max}}$. We use $\widetilde{\Pi}_\Omega$ to denote the orthogonal projection on the subspace $\widetilde{\mathcal{H}}_\Omega$.

To simulate the environment, we take $\widetilde{\rho}_{\rm E} = \ketbra{\widetilde{\psi}}{\widetilde{\psi}}$,
\begin{eqnarray}
\widetilde{H}_{\rm E} &=& - \sum_{\Omega} \Omega \widetilde{\Pi}_\Omega, \notag \\
\widetilde{B}_\beta &=& \sum_\omega \widetilde{b}_\beta(\omega).
\end{eqnarray}
On the quantum computer, we implement the time evolution with the Hamiltonian $\widetilde{H} = H_{\rm S} + \widetilde{H}_{\rm E} + \alpha \widetilde{H}_{\rm I}$ and the environment initial state $\widetilde{\rho}_{\rm E}$. Then TCL generator of the system evolution on the quantum computer $\widetilde{\mathcal{K}}$ is the same as the generator of the dynamics to be simulated $\mathcal{K}$ up to the $n$-th-order expansion, i.e.~$\widetilde{\KK}_m(t) = \widetilde{\PP} \KK_m(t)$ for all $m\leq n$, according to Theorem 1. The proof is given in Appendix.

\subsection{Discussion}

We can understand the algorithm as follows. By introducing the ancillary Hilbert space $\mathcal{H}_{\rm a}$, we can write the purification of the initial state $\rho_{\rm E}$ as $\ket{\psi} = \sum_\varepsilon \sqrt{p_\varepsilon} \ket{\Psi_\varepsilon}_{\rm E}\otimes\ket{\Phi_\varepsilon}_{\rm a}$, where $\ket{\Psi_\varepsilon}_{\rm E}$ is the eigenstate of the environment with the energy $\varepsilon$, and both $\{ \ket{\Psi_\varepsilon}_{\rm E} \}$ and $\{ \ket{\Phi_\varepsilon}_{\rm a} \}$ are orthonormal. Here, we have used that $\rho_{\rm E}$ is a stationary state. Then, we can write the Hamiltonian of the system, environment and ancillary system as $H' = H_{\rm S} + H_{\rm E}' + H_{\rm I}$, where $H_{\rm E}' = H_{\rm E} + H_{\rm a}$ and $H_{\rm a} = - \sum_\varepsilon \varepsilon \ketbra{\Phi_\varepsilon}{\Phi_\varepsilon}$. According to $H'$, the ancillary system is decoupled from the system and environment, and $\ket{\psi}$ is an eigenstate of $H_{\rm E}'$ with the energy $0$. Let $\Pi_{\rm E}$ be the orthogonal projection onto the relevant subspace $\mathcal{H}_{\rm E}^{\rm r}$, then $\widetilde{H} = \Pi_{\rm E} H' \Pi_{\rm E}$.

We would like to remark that, the ancillary system discussed here has been included in the environment $\widetilde{H}_{\rm E}$ on the quantum computer, which are not the ancillary qubits used for realising the evolution circuit shown in Fig.~\ref{fig:algorithm}(b).

Similar to the second-order simulation, the initial state of the environment on the quantum computer is always a pure state, and the pure state is not the ground state, because the frequency $\Omega$ can take both positive and negative values (see Fig.~\ref{fig:level}).

\section{Relaxation of the environment on the quantum computer}
\label{sec:reinitialisation}

Usually, higher-order terms of TCL equation are less significant, because of not only the weak coupling but also the huge energy and information capacity of the environment, i.e.~the influence of the system on the environment is small. However, on the quantum computer, the environment always has a finite size. As a result, high-order terms may become significant when the evolution time is long enough, specifically when the system and the environment exchange multiple excitations and the environment becomes saturate. Therefore, in this case we need to introduce the relaxation of the environment, i.e.~the dynamics implemented on the quantum computer is modified to $\frac{\partial}{\partial t} \rho(t) = -i[\widetilde{H}, \rho(t)] + \LL_{\rm R}\rho(t)$, where the Lindblad superoperator $\LL_{\rm R}$ acts on the environment and causes the relaxation. Evolution of such a Lindblad equation can also be implemented on the quantum computer~\cite{Bacon2001, Kliesch2011, Sweke2015, Candia2015, Sweke2016, Childs2017, Chenu2017}. In this section, we present three protocols for the environment relaxation.

Before we give relaxation protocols, we take the algorithm for the Markovian master equation simulation in Sec.~\ref{sec:Markovian} as an example to show the impact of the finite environment. According to $\widetilde{H}$, we have $\widetilde{\KK}_1(t) = \widetilde{\KK}_3(t) = 0$, and $\widetilde{\KK}_4(t)$ has four terms as shown in Eq.~(\ref{eq:K4}). The condition of the Markov approximation is the short correlation time $\tau_{\rm E}$ of the environment, i.e.~$\mean{ B_{\beta}(t) B_{\beta'}(t-s) }$ is insignificant if $s > \tau_{\rm E}$. Then, $\kk_2(t,t-s)$ is insignificant if $s > \tau_{\rm E}$. As a result, integrals of the last two terms in $\KK_4(t)$ leads to $\OO(\tau_{\rm E}^3)$. For example, the term $\kk_2(t,t_2)\kk_2(t_1,t_3)$ is significant only in the region defined by $t \geq t_1 \geq t_2$, $t \geq t_2 \geq t - \tau_{\rm E}$ and $t_2 \geq t_3 \geq t_1 - \tau_{\rm E}$. It is similar for $\kk_2(t,t_3)\kk_2(t_1,t_2)$. However, integrals of the second term result in $\OO(\tau_{\rm E}^2 t)$, because $\kk_2(t,t_1)\kk_2(t_2,t_3)$ is significant if $t-t_1 \leq \tau_{\rm E}$, $t_2-t_3 \leq \tau_{\rm E}$, but $t_1-t_2$ can be any value. Therefore, $\KK_4(t)$ is small only if the second term and the first term $\kk_4(t,t_1,t_2,t_3)$ cancel with each other, i.e.~$\kk_4(t,t_1,t_2,t_3) \simeq \kk_2(t,t_1)\kk_2(t_2,t_3)$ when $t_1-t_2>\tau_{\rm E}$, which means that two excitations in the environment do not interfere with each other if they are separated by a time interval bigger than $\tau_{\rm E}$. However, in our algorithm for simulating the second-order equation, at most only one excitation can exist in the environment on the quantum computer, and the first excitation always prevents the second excitation, therefore $\widetilde{\KK}_4(t) = \OO(\tau_{\rm E}^2 t)$.

As an example, we consider one of sixteen of terms in $\widetilde{\kk}_4(t,t_1,t_2,t_3)$,
\begin{eqnarray}
&& \Tr[\widetilde{B}_{\beta}(t)\widetilde{B}_{\beta_3}(t_3)\widetilde{\rho}_{\rm E}\widetilde{B}_{\beta_2}(t_2)\widetilde{B}_{\beta_1}(t_1)] \notag \\
&\times & A_{\beta}(t)A_{\beta_3}(t_3) \bullet A_{\beta_2}(t_2)A_{\beta_1}(t_1).
\end{eqnarray}
Because at most only one excitation can exist in the environment, the contribution of the following components is nonzero [see Eq.~(\ref{eq:B2nd})]: the $\sigma_{l}^\dag(\omega)$ component of $\widetilde{B}_{\beta_3}$, the $\sigma_{l}(\omega)$ component of $\widetilde{B}_{\beta_2}$, the $\sigma_{l}^\dag(\omega)$ component of $\widetilde{B}_{\beta_1}$ and the $\sigma_{l}(\omega)$ component of $\widetilde{B}_{\beta}$. As a result,
\begin{eqnarray}
&& \Tr[\widetilde{B}_{\beta}(t)\widetilde{B}_{\beta_3}(t_3)\widetilde{\rho}_{\rm E}\widetilde{B}_{\beta_2}(t_2)\widetilde{B}_{\beta_1}(t_1)] \notag \\
&=& \mean{ B_{\beta}(t) B_{\beta_3}(t_3) }\times\mean{ B_{\beta_1}(t_1) B_{\beta_2}(t_2) }^*,
\end{eqnarray}
which is significant if $t-t_3 \leq \tau_{\rm E}$. We remark that $t_1$ and $t_2$ are between $t$ and $t_3$. The corresponding term in $\widetilde{\kk}_2(t,t_1)\widetilde{\kk}_2(t_2,t_3)$ is
\begin{eqnarray}
&& \Tr[\widetilde{B}_{\beta}(t)\widetilde{\rho}_{\rm E}\widetilde{B}_{\beta_1}(t_1)]\times\Tr[\widetilde{B}_{\beta_3}(t_3)\widetilde{\rho}_{\rm E}\widetilde{B}_{\beta_2}(t_2)] \notag \\
&\times & A_{\beta}(t)A_{\beta_3}(t_3) \bullet A_{\beta_2}(t_2)A_{\beta_1}(t_1),
\end{eqnarray}
where
\begin{eqnarray}
&& \Tr[\widetilde{B}_{\beta}(t)\widetilde{\rho}_{\rm E}\widetilde{B}_{\beta_1}(t_1)]\times\Tr[\widetilde{B}_{\beta_3}(t_3)\widetilde{\rho}_{\rm E}\widetilde{B}_{\beta_2}(t_2)] \notag \\
&=& \mean{ B_{\beta}(t) B_{\beta_1}(t_1) }^*\times\mean{ B_{\beta_2}(t_2) B_{\beta_3}(t_3) }.
\end{eqnarray}
For any value of $t_1-t_2$, the corresponding term in $\widetilde{\kk}_2(t,t_1)\widetilde{\kk}_2(t_2,t_3)$ can be significant. Therefore, $\widetilde{\kk}_4(t,t_1,t_2,t_3) \simeq \widetilde{\kk}_2(t,t_1)\widetilde{\kk}_2(t_2,t_3)$ does not hold when $t_1-t_2>\tau_{\rm E}$.

Next, we show that $\widetilde{\KK}_4(t)$ can be suppressed by introducing the environment relaxation.

\subsection{Reinitialisation protocol}

A way to realise the environment relaxation is the periodic reinitialisation of the environment state at time $j \tau$, where $\tau$ is the period, and $j$ is an integer~\cite{Terhal2000, Wang2011}. In such a protocol, correlation functions on the quantum computer with $m\leq n$ are significantly modified by the relaxation and become
\begin{eqnarray}
&& \Tr \big[ \widetilde{\BB}_{\beta}(t,\nu) \cdots \widetilde{\BB}_{\beta_{m-1}}(t_{m-1},\nu_{m-1}) \widetilde{\rho}_{\rm E} \big] \notag \\
&=& \Tr \big[ \BB_{\beta}(t,\nu) \VV_1 \cdots \VV_{m-1} \BB_{\beta_{m-1}}(t_{m-1},\nu_{m-1}) \rho_{\rm E} \big],~~~
\label{eq:VV}
\end{eqnarray}
where $\VV_i = [\openone]$ if $t_{i-1}$ ($t_0 = t$) and $t_i$ are in the same period, i.e.~$(j+1)\tau > t_{i-1} \geq t_i > j\tau$ for any integer $j$, otherwise $\VV_i = \PP$. Here, $\PP$ is the projection onto the state $\rho_{\rm E}$, $\openone$ is the identity operator, and $[U]\rho = U \rho U^\dag$.

For two-time correlation functions, $\mean{ \widetilde{B}_{\beta}(t) \widetilde{B}_{\beta'}(s) } = \mean{ B_{\beta}(t) B_{\beta'}(s) }$ only if $t$ and $s$ are in the same period, otherwise it is zero. We note that even if $t$ and $s$ are close, the two-time correlation function is zero if they are in different periods. Because of the reinitialisation, $\widetilde{\kk}_4(t,t_1,t_2,t_3) = \widetilde{\kk}_2(t,t_1)\widetilde{\kk}_2(t_2,t_3)$ if $t_1-t_2>\tau$, therefore the fourth order term is suppressed to $\widetilde{\KK}_4(t) = \OO(\tau_{\rm E}^2 \tau)$.

\subsection{Projective dissipation protocol}

We can implement the environment reinitialisation stochastically at a constant rate of $\Gamma$, i.e.~the corresponding Lindblad superoperator is $\LL_{\rm R} = \Gamma (\widetilde{\PP} - [\openone])$. With such a relaxation term, correlation functions on the quantum computer with $m\leq n$ can also be expressed in the form of Eq.~(\ref{eq:VV}), but $\VV_i = e^{-\Gamma s_i}[\openone] + (1-e^{-\Gamma s_i})\PP$, where $s_i = t_{i-1}-t_i$. We remark that, $\LL_{\rm R}$ and the environment Hamiltonian $-i[\widetilde{H}_{\rm E}, \bullet]$ are commutative, because $\widetilde{\rho}_E$ is a stationary state.

Using the projective dissipation protocol, two-time correlation functions become $\mean{ \widetilde{B}_{\beta}(t) \widetilde{B}_{\beta'}(s) } = e^{-\Gamma(t-s)} \mean{ B_{\beta}(t) B_{\beta'}(s) }$.

In some cases, correlation functions can be exactly reproduced on the quantum computer even with the presence of environment dissipation $\LL_{\rm R}$. For the second-order equation simulation, if Fourier transformations of $e^{\Gamma(t-s)} \mean{ B_{\beta}(t) B_{\beta'}(s) }$ yield a set of positive matrices $\gamma(\omega)$, we can choose coefficients $g_{\beta, l}(\omega)$ according to $e^{\Gamma(t-s)} \mean{ B_{\beta}(t) B_{\beta'}(s) }$, so that $\mean{ \widetilde{B}_{\beta}(t) \widetilde{B}_{\beta'}(s) } = \mean{ B_{\beta}(t) B_{\beta'}(s) }$ when the dissipation is introduced. It is similar for higher-order equation simulations.

If correlation functions cannot be exactly reproduced, we may need to take $\Gamma \ll \tau_{\rm E}^{-1}$, so that correlation functions are not significantly modified. The relaxation time of the environment is $\Gamma^{-1}$. Therefore excitations in the environment do not interfere with each other if they are separated by a time interval bigger than $\Gamma^{-1}$, i.e.~$\widetilde{\KK}_4(t) = \OO(\tau_{\rm E}^2 \Gamma^{-1})$ in the second-order equation simulation.

\subsection{Conditional projective dissipation protocol}
\label{sec:CPDP}{

An optimal dissipation protocol relaxes the environment as soon as possible but does not modify correlations functions. Here we present such an environment dissipation protocol motivated by a typical open quantum system, an atom coupled to the free space in the vacuum state as shown in Fig.~\ref{fig:photon}(a). The excited state of the atom decays into the ground state by emitting a photon into the free space. Because once the photon is emitted it leaves the atom and never comes back, the decay is irreversible. The correlation time depends on the length of the photon wave-package, because once the wave-package is out of the reach of the coupling, the photon cannot affect the atom anymore. Therefore, if a black material is placed at a finite but sufficient distance from the atom and absorbs photons [see Fig.~\ref{fig:photon}(b)], the black material does not affect the evolution of the atom (neglecting the radiation from the material). We are interested in cases that the correlation time is short compared with the coupling between the system and environment, so that the expansion of TCL master equation is reasonable. We find that when the correlation time is short, only a subspace of the correlation-relevant state space $\mathcal{H}_{\rm E}^{\rm r}$ effectively contributes to correlation functions. Therefore, we can let the environment evolve without dissipation within the subspace, i.e.~the left side of the black material, and the environment dissipates once its state is out of the subspace, i.e.~the right side of the black material.

\begin{figure}[tbp]
\centering
\includegraphics[width=1\linewidth]{\figpath 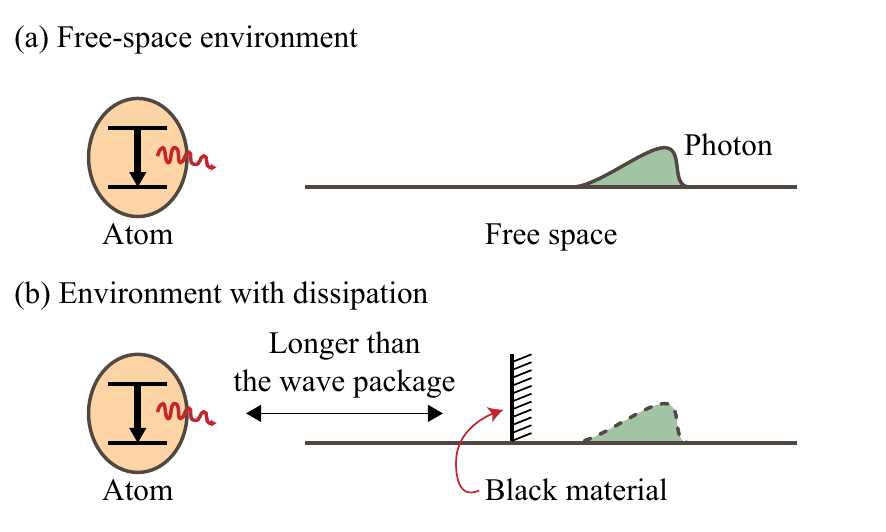}
\caption{
An example of dissipation caused by the environment. (a) An atom coupled to the free space in the vacuum state. (b) A black material is placed at a distance from the atom and absorbs photons.
}
\label{fig:photon}
\end{figure}

First, we consider the second-order equation simulation. We will generalise the protocol to higher-order equation simulations later. For the second-order equation simulation, we show that using the conditional dissipation protocol, the environment relaxes in the time scale $\sim \tau_{\rm E}$, but two-time correlation functions are only slightly modified.

Suppose that the environment spectrum is discretised with the uniform spacing $\delta \omega$, then each frequency $\omega$ corresponds to an integer $k$ and $\omega = k \delta\omega$. We apply the Fourier transformation to states $\ket{\omega, l}$ and define $\ket{x, l} \equiv \frac{1}{\sqrt{N_\omega}}\sum_\omega e^{-i\frac{\omega x}{c}} \ket{\omega, l}$, where $c = \frac{N_\omega \delta_\omega}{2\pi}$ and $x = 0, \ldots, N_\omega-1$. These states form a ring as shown in Fig.~\ref{fig:ring}. For a wave-package in the from $\sum_x a_x\ket{x,l}$, the evolution driven by $\widetilde{H}_{\rm E}$ transports the wave-package along the ring [see Fig.~\ref{fig:ring}(a)], i.e.~$e^{-i\widetilde{H}_{\rm E}t}(\sum_x a_x\ket{x,l}) = \sum_x a_x\ket{x+ct,l}$ when $ct$ is an integer. Here $\ket{x+N_\omega} \equiv \ket{x}$. Therefore, the evolution is periodic, and the period is $\frac{N_\omega}{c} = \frac{2\pi}{\delta \omega}$.

We would like to note that using the uniformly discretised spectrum on the quantum computer, two-time correlation functions with $s = t-t_1$ in the interval $[0, \frac{2\pi}{\delta \omega}]$ are reproduced in the form of Fourier series, which converges as $N_\omega \rightarrow \infty$. The optimal range of $k$ depends on correlation functions. Without loss of generality, we suppose $N_\omega$ is odd, and we take $k = -\frac{N_\omega-1}{2},\ldots,0,1,\ldots,\frac{N_\omega-1}{2}$.

\begin{figure}[tbp]
\centering
\includegraphics[width=1\linewidth]{\figpath 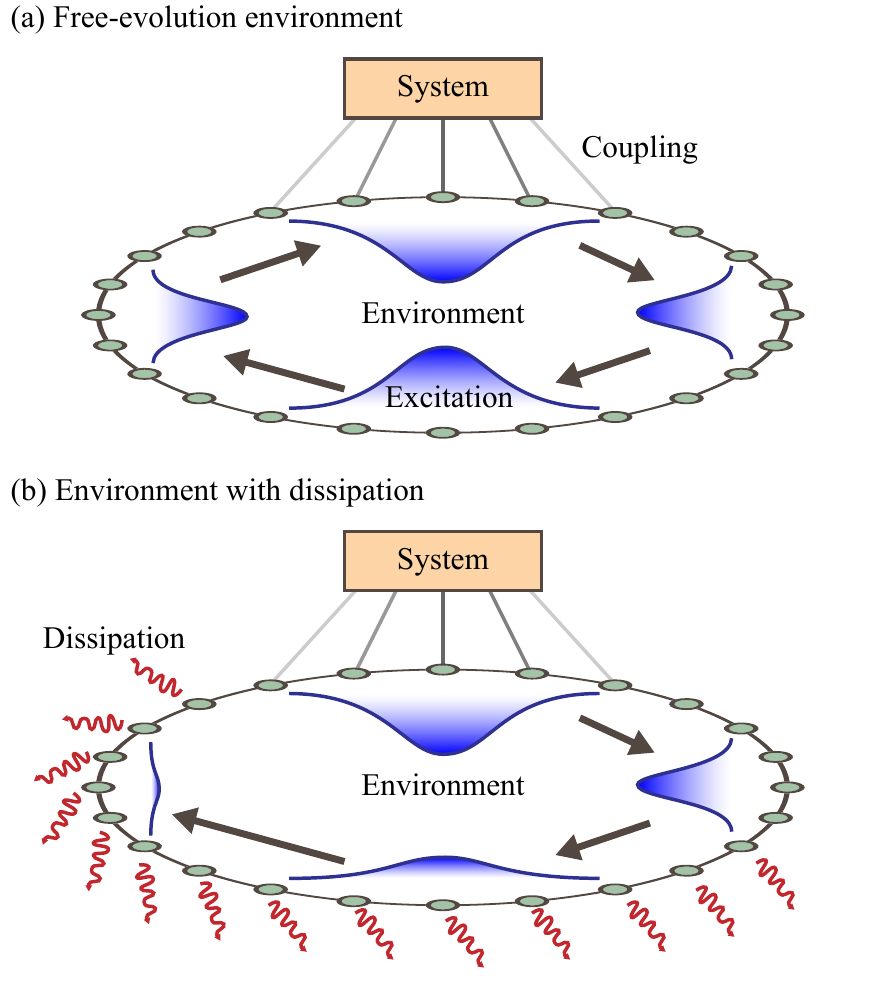}
\caption{
Simulation of the second-order equation. (a) The system is coupled to the environment via local interaction. Without dissipation, an excitation in the environment leaves the interaction region but never disappears. (b) The dissipation is switched on in the region without interaction. The excitation disappears when it leaves the interaction region.
}
\label{fig:ring}
\end{figure}

In the $x$ representation, we can re-express interaction operators as
\begin{eqnarray}
\widetilde{B}_\beta = \sum_{x,l} g_{\beta,l}(x) \sigma_l(x) + {\rm h.c.},
\end{eqnarray}
where $\sigma_l(x) \equiv \ketbra{v}{x,l}$ and
\begin{eqnarray}
g_{\beta,l}(x) = \frac{1}{\sqrt{N_\omega}}\sum_\omega e^{-i\frac{2\pi k x}{N_\omega}} g_{\beta,l}(\omega).
\end{eqnarray}
Therefore, $\widetilde{B}_\beta \ket{v} = \sum_{x,l} g_{\beta,l}^*(x) \ket{x,l}$ is a wave-package in the $x$ space. Without the dissipation, the correlation function
\begin{eqnarray}
\mean{\widetilde{B}_{\beta}(t) \widetilde{B}_{\beta'}(t-s)} = \bra{\rm v} \widetilde{B}_{\beta} e^{-i \widetilde{H}_{\rm E} s} \widetilde{B}_{\beta'} \ket{\rm v}
\end{eqnarray}
is the overlap between two wavepackages $\widetilde{B}_\beta \ket{v}$ and $e^{-i \widetilde{H}_{\rm E} s} \widetilde{B}_{\beta'} \ket{\rm v}$.

If $\beta = \beta'$, the correlation function is maximised at $s = 0$. The second wavepackage moves in the $x$ space with the speed $c$ without dispersion. As a result, the correlation function decreases with the time $s$. The correlation function vanishes at $s \sim \tau_{\rm E}$, which implies that the wavepackage is localised in the $x$ space with the width $~x_{\rm E} \equiv c\tau_{\rm E}$. Because the wavepackage is created by the coupling, the coupling strength $g_{\beta,l}(x)$ is also localised in the $x$ space with the same width as shown in Fig.~\ref{fig:ring}. The localised coupling means that the matrix $\gamma(\omega)$ varies slowly with the frequency $\omega$. In the following, we assume that the coupling is localised in the region $0 \leq x \leq x_{\rm E}$, which is reasonable when the system is coupled to the environment via local interactions.

The conditional dissipation protocol works as follows. In the region $0 \leq x \leq x_{\rm T}$, where $x_{\rm T} \geq x_{\rm E}$, a wavepackage propagates freely without dissipation, such that two-time correlation functions can be reproduced. We remark that two-time correlation functions are only determined by the wavepackage in the region $0 \leq x \leq x_{\rm E}$. In the region $x > x_{\rm T}$, the excitation decays at the rate of $\Gamma$, and the environment is stochastically reinitialised to the state $\ket{\rm v}$.

To implement the conditional dissipation, at the rate of $\Gamma$ we perform a measurement to find out whether the environment is in states with $x > x_{\rm T}$, i.e.~the projection $\Pi = \sum_{x > x_{\rm T}, l} \sigma_l(x)^\dag \sigma_l(x)$. The environment reinitialisation is implemented depending on the measurement outcome. The corresponding Lindblad superoperator reads $\mathcal{L}_{\rm R} = \Gamma (\widetilde{\PP}[\Pi]+[\openone-\Pi]-[\openone])$. Because of the dissipation, the wavepackage disappears before the revival. When the wavepackage disappears, the environment is reinitialised, and the next excitation can enter the environment.

The dissipation may cause quantum Zeno effect, which can prevent the wavepackage from entering the dissipation region $x > x_{\rm T}$. The propagation from $\ket{x}$ to $\ket{x+1}$ takes the time $c^{-1} = \frac{2\pi}{N_\omega \delta\omega}$. Therefore, the quantum Zeno effect is weak if $\Gamma \ll c$. Here, $c^{-1}$ corresponds to the time resolution of the environment. When the time resolution is fine, we should have $\tau_{\rm E} \gg c^{-1}$. In this case, we can take $\Gamma = \tau_{\rm E}^{-1}$ and $x_{\rm T} = x_{\rm E}$, such that excitations in the environment do not interfere with each other if they are separated by a time interval bigger than $\tau_{\rm E}$, i.e.~$\KK_4(t) = \OO(\tau_{\rm E}^3)$.

As an example, let us consider a simple case that the interaction Hamiltonian of the system and environment consists of only one term, reading $\widetilde H_{\rm I}=\alpha A \otimes \widetilde{B}$, with
\begin{equation}
\widetilde{B} = \sum_\omega \left[ g(\omega) \ketbra{\omega}{{\rm v}} + g^*(\omega) \ketbra{{\rm v}}{\omega} \right],
\label{eq:B}
\end{equation}
where the environment spectrum is discretised with the uniform spacing $\delta \omega$, and $g(\omega) = a \sqrt{\frac{2\bar\gamma}{\omega^2+{\bar\gamma}^2} \frac{\delta\omega}{2\pi}}$, where $a$ and $\bar\gamma$ are constants with the dimensions of frequency. With such an environment, the reconstructed correlation function is $\langle \widetilde{B}(t) \widetilde{B}(t-s) \rangle = a^2 e^{-\bar\gamma \abs{s}}$ in the limit $N_\omega \rightarrow \infty$. In Fig.~\ref{fig:wavepackage}, we plot the wavepackage $\widetilde{B} \ket{\rm v}$ in the $x$ space with the conditional dissipation. One can see that the wavepackage travels freely from left to right, until it enters the dissipation zone in which it quickly diminishes.

{\bf Conditional reinitialisation.} To avoid the quantum Zeno effect, we can replace the continuous-time dissipation with periodic conditional reinitialisation. The Lindblad superoperator becomes time-dependent and reads $\mathcal{L}_{\rm R} = \Gamma(t) (\widetilde{\PP}[\Pi]+[\openone-\Pi]-[\openone])$, where $\Gamma(t) = +\infty$ when $t = j c^{-1}$, $\Gamma(t) = 0$ when $t \neq j c^{-1}$, and $j$ is an integer. In other words, a wavepackage can propagate freely in each time interval with the length $c^{-1}$, i.e.~the wavepackage can propagate by one site in the $x$ space, and at each time $t = j c^{-1}$ the environment is conditionally reinitialised by implementing the operation $\widetilde{\PP}[\Pi]+[\openone-\Pi]$. In this way, without affecting the propagation of the wavepackage in the region $x \leq x_{\rm E}$, the environment relaxes in the time scale $\tau_{\rm E}$. Here we take $x_{\rm T} = x_{\rm E}$.

\begin{figure}[tbp]
\centering
\includegraphics[width=1\linewidth]{\figpath 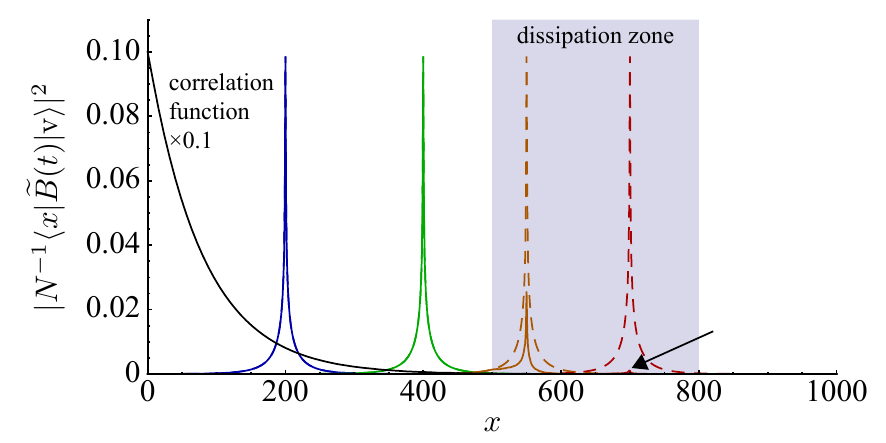}
\caption{
Numerical results for the wavepackage $N^{-1}\widetilde{B}(t) \ket{\rm v}$, where $N = \norm{\widetilde{B}(0) \ket{\rm v}}$, i.e.~the wavepackage is normalised at $t = 0$. The probability in the state $\ket{x}$ is plotted. We take $N_\omega = 1001$, $\delta\omega = 0.001$, $\Gamma=0.004$ and $\bar\gamma=0.002$. The conditional dissipation is introduced between $x=500$ and $x=800$. The wavepackage initialises at $x=0$ when $t=0$. From left to right, the blue, green, yellow and red solid curves correspond to time $t = (200, 400, 550, 700) \times 2\pi/N_\omega\delta\omega$, respectively. The wavepackage is computed using the quantum trajectory approach~\cite{Gardiner2000, Carmichael1993}, and each curve is obtained with $1000$ instances. For comparison, dashed curves denote the wavepackage when the dissipation is turned off. The black curve represents the correlation function $e^{-\bar\gamma \abs{s}}$ with $s = x/c$, which vanishes at about $x = 300$. Therefore, the wavepackage centered at $x > 300$ does not contribute to the correlation function. The wavepackage vanishes (see the arrow) after it enters the dissipation zone (marked in gray).
}
\label{fig:wavepackage}
\end{figure}

\subsubsection*{Generalisation to the higher-order simulations}

In this section, we discuss how to generalise the conditional dissipation (or reinitialisation) protocol to higher-order simulations. For environments similar to the case in Fig.~\ref{fig:photon}, the system only affects the part of the environment close to it, its influence (excitations) propagates in the environment, and the part close to the system relaxes in the time scale $\tau_{\rm E}$, i.e.~correlations
\begin{eqnarray}
\Delta &=& \Tr \big[ \cdots \BB_{\beta_{i}}(t_{i},\nu_{i}) \BB_{\beta_{i+1}}(t_{i+1},\nu_{i+1}) \cdots \rho_{\rm E} \big] \notag \\
&&- \Tr \big[ \cdots \BB_{\beta_{i}}(t_{i},\nu_{i}) \PP \BB_{\beta_{i+1}}(t_{i+1},\nu_{i+1}) \cdots \rho_{\rm E} \big] ~~~
\label{eq:Delta}
\end{eqnarray}
are negligible when $t_{i} - t_{i+1} > \tau_{\rm E}$. We remark that because correlation functions are reproduced, it is the same for the environment on the quantum computer. For such environments, the reinitialisation operation $\PP$ on $[e^{-iH_{\rm E}s}] \BB_{\beta_{i+1}}(t_{i+1},\nu_{i+1}) \cdots \rho_{\rm E}$ does not affect correlation functions. Here $s > \tau_{\rm E}$, and $[e^{-iH_{\rm E}s}]$ is the superoperator denoting the free evolution of the environment. To implement the condition dissipation, we need to find the proper projection $\Pi$ representing the space of states that the influence of the system has left the interaction region.

As an example, we consider the fourth-order simulation using the environment on the quantum computer given by
\begin{eqnarray}
\widetilde{H}_{\rm E} &=& \sum_{\omega} \omega \ketbra{\omega}{\omega} \notag \\
&&+ \sum_{\omega_1,\omega_2} (\omega_1+\omega_2) \ketbra{\omega_1,\omega_2}{\omega_1,\omega_2}
\label{eq:fourth-order_H}
\end{eqnarray}
and
\begin{eqnarray}
\widetilde{B}_\beta &=& \sum_{\omega} g_{\beta,\omega} \ketbra{\rm v}{\omega} \notag \\
&&+ \sum_{\omega,\omega_1,\omega_2} g_{\beta,\omega,\omega_1,\omega_2} \ketbra{\omega}{\omega_1,\omega_2} + {\rm h.c.}
\label{eq:fourth-order_B}
\end{eqnarray}
Here, $\ket{\rm v}$ denotes the vacuum state and the initial state of the environment, i.e.~$\widetilde{\rho}_{\rm E} = \ketbra{\rm v}{\rm v}$, $\ket{\omega}$ denotes the state of one excitation with the frequency $\omega$, and $\ket{\omega_1, \omega_2}$ denotes the state of two excitations with frequencies $\omega_1$ and $\omega_2$, respectively. By choosing coupling coefficients $g_{\beta,\omega}$ and $g_{\beta,\omega,\omega_1,\omega_2}$, we can reproduce some reservoir correlation functions (see Appendix). The general algorithm for higher-order simulations is given in Sec.~\ref{sec:algorithm}.

Correlation functions reproduced in the environment given by Eq.~(\ref{eq:fourth-order_H}) and Eq.~(\ref{eq:fourth-order_B}) are
\begin{eqnarray}
\bra{\rm v} \widetilde{B}_\beta(s) \widetilde{B}_{\beta_1}(s_1) \ket{\rm v} = \sum_\omega e^{-i\omega(s-s_1)} g_{\beta,\omega} g_{\beta_1,\omega}^*,
\end{eqnarray}
\begin{eqnarray}
\bra{\rm v} \widetilde{B}_\beta(s) \widetilde{B}_{\beta_1}(s_1) \widetilde{B}_{\beta_2}(s_2) \ket{\rm v} = 0
\end{eqnarray}
and
\begin{eqnarray}
&& \bra{\rm v} \widetilde{B}_\beta(s) \widetilde{B}_{\beta_1}(s_1) \widetilde{B}_{\beta_2}(s_2) \widetilde{B}_{\beta_3}(s_3) \ket{\rm v} \notag \\
&=& \bra{\rm v} \widetilde{B}_\beta(s) \widetilde{B}_{\beta_1}(s_1) \ket{\rm v} \bra{\rm v} \widetilde{B}_{\beta_2}(s_2) \widetilde{B}_{\beta_3}(s_3) \ket{\rm v} \notag \\
&&+ \sum_{\omega,\omega',\omega_1,\omega_2} e^{-i\omega'(s-s_1)} e^{-i(\omega_1+\omega_2)(s_1-s_2)} e^{-i\omega(s_2-s_3)} \notag \\
&&\times g_{\beta,\omega'} g_{\beta_1,\omega',\omega_1,\omega_2} g_{\beta_2,\omega,\omega_1,\omega_2}^* g_{\beta_3,\omega}^*.
\end{eqnarray}

Similar to the second-order simulation, we suppose that the environment is discretised with the uniform spacing $\delta$, i.e.~$\omega,\omega_1,\omega_2 = k\delta\omega$ and $k = -\frac{N_\omega-1}{2},\ldots,0,1,\ldots,\frac{N_\omega-1}{2}$. Then, by applying the Fourier transformation, we have
\begin{eqnarray}
\ket{x} &\equiv & \frac{1}{\sqrt{N_\omega}}\sum_\omega e^{-i\frac{2\pi \omega x}{N_\omega \delta_\omega}} \ket{\omega}, \\
\ket{x_1,x_2} &\equiv & \frac{1}{N_\omega}\sum_{\omega_1,\omega_2} e^{-i\frac{2\pi (\omega_1 x_1 + \omega_2 x_2)}{N_\omega \delta_\omega}} \ket{\omega_1,\omega_2}.
\end{eqnarray}
In the $x$ representation,
\begin{eqnarray}
\widetilde{B}_\beta &=& \sum_{x} g_{\beta,x} \ketbra{{\rm v}}{x} \notag \\
&&+ \sum_{x,x_1,x_2} g_{\beta,x,x_1,x_2} \ketbra{x}{x_1,x_2} + {\rm h.c.},
\end{eqnarray}
where
\begin{eqnarray}
g_{\beta,x} &=& \frac{1}{\sqrt{N_\omega}}\sum_\omega e^{-i\frac{2\pi \omega x}{N_\omega \delta_\omega}} g_{\beta,\omega}, \notag \\
g_{\beta,x,x_1,x_2} &=& \frac{1}{N_\omega^{\frac{3}{2}}}\sum_{\omega,\omega_1,\omega_2} e^{i\frac{2\pi (\omega x - \omega_1 x_1 - \omega_2 x_2)}{N_\omega \delta_\omega}} g_{\beta,\omega,\omega_1,\omega_2}.~~~
\end{eqnarray}
If we only consider correlation functions at discretised times, i.e.~$cs,cs_1,cs_2,cs_3$ are integers, two-time and four-time correlation functions are
\begin{eqnarray}
\bra{\rm v} \widetilde{B}_\beta(s) \widetilde{B}_{\beta_1}(s_1) \ket{\rm v} = \sum_x g_{\beta,x+c(s-s_1)} g_{\beta_1,x}^*,
\end{eqnarray}
and
\begin{eqnarray}
&& \bra{\rm v} \widetilde{B}_\beta(s) \widetilde{B}_{\beta_1}(s_1) \widetilde{B}_{\beta_2}(s_2) \widetilde{B}_{\beta_3}(s_3) \ket{\rm v} \notag \\
&=& \bra{\rm v} \widetilde{B}_\beta(s) \widetilde{B}_{\beta_1}(s_1) \ket{\rm v} \bra{\rm v} \widetilde{B}_{\beta_2}(s_2) \widetilde{B}_{\beta_3}(s_3) \ket{\rm v} \notag \\
&&+ \sum_{x,x',x_1,x_2} g_{\beta,x'+c(s-s_1)} g_{\beta_1,x',x_1+c(s_1-s_2),x_2+c(s_1-s_2)} \notag \\
&&\times g_{\beta_2,x+c(s_2-s_3),x_1,x_2}^* g_{\beta_3,x}^*.
\end{eqnarray}
Three-time correlation functions are zero.

Correlations in Eq.~(\ref{eq:Delta}) are negligible if the system is only coupled to environment states $\ket{x}$ and $\ket{x_1,x_2}$ with $x,x_2 = 0,\ldots,x_{\rm E}$, where $x_{\rm E} = c\tau_{\rm E}$. We have $\Delta = 0$ if $g_{\beta,x} = 0$ when $x > x_{\rm E}$, $g_{\beta,x,x_1,x_2} = 0$ when $x_2 > x_{\rm E}$, and $g_{\beta,x,x_1,x_2} = \delta_{x,x_1}g_{\beta,x_2}$ when $x,x_1 > x_{\rm E}$. It is obvious for two-time correlation functions. For four-time correlation functions, considering values of $\nu$ [see Eq.~(\ref{eq:corr_fun})], there are $16$ of them, but only four of them are independent, which are
\begin{eqnarray}
&& \bra{\rm v} \widetilde{B}_\beta(t) \widetilde{B}_{\beta_1}(t_1) \widetilde{B}_{\beta_2}(t_2) \widetilde{B}_{\beta_3}(t_3) \ket{\rm v}, \notag \\
&& \bra{\rm v} \widetilde{B}_{\beta_1}(t_1) \widetilde{B}_\beta(t) \widetilde{B}_{\beta_2}(t_2) \widetilde{B}_{\beta_3}(t_3) \ket{\rm v}, \notag \\
&& \bra{\rm v} \widetilde{B}_{\beta_2}(t_2) \widetilde{B}_\beta(t) \widetilde{B}_{\beta_1}(t_1) \widetilde{B}_{\beta_3}(t_3) \ket{\rm v}, \notag \\
&& \bra{\rm v} \widetilde{B}_{\beta_2}(t_2) \widetilde{B}_{\beta_1}(t_1) \widetilde{B}_\beta(t) \widetilde{B}_{\beta_3}(t_3) \ket{\rm v},
\end{eqnarray}
where $t \geq t_1 \geq t_2 \geq t_3$. We can check that $\Delta = 0$ for all of them. Here, we have assumed that $N_\omega \gg x_{\rm E}$. Therefore, to implement the conditional dissipation, we can take the projection $\Pi = \sum_{x>x_{\rm E}} \ketbra{x}{x} + \sum_{x_1}\sum_{x_2>x_{\rm E}} \ketbra{x_1,x_2}{x_1,x_2}$.

\section{Circuit implementation, time cost and hardware resource requirement}
\label{sec:cost}

Given the Hamiltonian $\widetilde{H}$, an initial state of the system $\rho_{\rm S}(0)$, and the initial state of the environment $\widetilde{\rho}_{\rm E}$, we can implement the unitary dynamics $\rho(t) = e^{-i \widetilde{H}t} \rho_{\rm S}(0) \otimes \widetilde{\rho}_{\rm E} e^{i \widetilde{H}t}$ on the quantum computer. Then, $\rho(t)$ is a solution of the evolution equation $\frac{\partial}{\partial t} \widetilde{\PP}\rho(t) = \widetilde{\KK}(t)\widetilde{\PP}\rho(t)$. According to discussions in the previous section, $\widetilde{\KK}(t)$ and $\KK(t)$ are the same up to the $n$-th-order expansion.

We can implement the dynamics of $\widetilde{H}$ using the Trotterisation algoirthm~\cite{Lloyd1996}. Let $N_{\rm S}$ be the number of qubits representing the system, then the total number of qubits used in the simulation is $N_{\rm S} + N_{\rm E}$. System operators can always be expanded using Pauli operators as $H_{\rm S} = \sum_{\sigma \in S_H} f_{H,\sigma} \sigma$ and $A_\beta = \sum_{\sigma \in S_\beta} f_{\beta, \sigma} \sigma$. Here, $S_H$ and $S_\beta$ are subsets of $N_{\rm S}$-qubit Pauli operators. Similarly, environment operators can also be expanded using Pauli operators as $\widetilde{H}_{\rm E} = \sum_{\sigma \in E_H} h_{H,\sigma} \sigma$ and $\widetilde{B}_\beta = \sum_{\sigma \in E_\beta} h_{\beta, \sigma} \sigma$. Here, $E_H$ and $E_\beta$ are subsets of $N_{\rm E}$-qubit Pauli operators. Expansion coefficients $f$ and $h$ are all real, because these expanded operators are all Hermitian. Using Trotterisation, the evolution implemented on the quantum computer is
\begin{eqnarray}
U_{N_{\rm T}} &=& \prod_{i = 1}^{N_{\rm T}} \left[ \left( \prod_{\sigma \in S_H} e^{-i \sigma \frac{f_{H,\sigma} t}{N_{\rm T}}} \right)
\otimes \left( \prod_{\sigma' \in E_H} e^{-i \sigma' \frac{h_{H,\sigma'} t}{N_{\rm T}}} \right) \right] \notag \\
&&\times
\left( \prod_{\beta} \prod_{\sigma \in S_\beta} \prod_{\sigma' \in E_\beta} e^{-i \sigma \otimes \sigma' \frac{\alpha f_{\beta,\sigma} h_{\beta,\tau} t}{N_{\rm T}}} \right),
\end{eqnarray}
where $N_{\rm T}$ is the number of Trotter steps, and each exponential of $k$-qubit Pauli operator can be implemented on the quantum computer with up to $2(k-1)$ CNOT gates and $2k+1$ single-qubit gates~\cite{Whitfield2011}. Therefore, the total number of gates $N_{\rm G}$ is less than $[(4N_{\rm S}-1)\abs{S_H} + (4N_{\rm E}-1)\abs{E_H} + (4N_{\rm S} + 4N_{\rm E} -1) \sum_{\beta} \abs{S_\beta} \abs{E_\beta}] N_{\rm T}$.

The Trotter-Suzuki decomposition is approximate, and the difference between $U_{N_{\rm T}}$ and $e^{-i \widetilde{H}}$ is
\begin{eqnarray}
\epsilon_{\rm Trotter} = \norm{U_{N_{\rm T}} - e^{-i \widetilde{H}}} \sim \frac{N_{\rm terms} \norm{\widetilde{H}}^2 t^2}{N_{\rm T}},
\end{eqnarray}
where $\norm{\bullet}$ denotes the operator norm, and $N_{\rm terms} = \abs{S_H} + \abs{E_H} + \sum_{\beta} \abs{S_\beta} \abs{E_\beta}$ is the number of terms in the Hamiltonian. We can prove that $\norm{\widetilde{B}_\beta} \leq \norm{B_\beta}$ (see Appendix), therefore the norm of the Hamiltonian has the upper bound
\begin{eqnarray}
\norm{\widetilde{H}} \leq \norm{H_{\rm S}} + n\times\max\{\abs{\omega}\} + \alpha \sum_{\beta} \norm{A_\beta}\norm{B_\beta}.~~~~~
\end{eqnarray}
Here, $\max\{\abs{\omega}\} \sim \norm{H_{\rm E}}$. However, usually it is sufficient to truncate the frequency at $\max\{\abs{\omega}\} \sim \norm{H_{\rm S}}$ when the coupling is weak.

Usually, for a Hamiltonian with local interactions, the number of terms in the Hamiltonian, i.e.~each of $\abs{S_H}$, $\abs{S_\beta}$ and $N_\beta$, is a polynomial with respect to the system size $N_{\rm S}$.

The number of qubits required for simulating the environment is $N_{\rm E} \sim \lfloor \frac{n}{2} \rfloor \log_2 (N_\omega N_\beta)$, because $d_{\rm E} \sim (N_\omega N_\beta)^{\lfloor \frac{n}{2} \rfloor}$. According to the maximum number of environment Pauli operators, we have $\abs{E_H},\abs{E_\beta} \sim 4^{N_{\rm E}} \sim (N_\omega N_\beta)^n$.

To implement the conditional dissipation, we may need to introduce only one more qubit for the measurement of $\Pi$, i.e.~we can use the state $\ket{1}$ of the qubit to indicate the subspace. Because the conditional dissipation operation is performed at a low rate, the cost of gate number is small compared with the unitary evolution.

In summary, the simulation requires $N_{\rm E} \sim \lfloor \frac{n}{2} \rfloor \log_2 (N_\omega N_\beta)$ qubits to simulate the environment. The number of terms in the Hamiltonian is $N_{\rm terms} = \OO(N_\omega^n N_\beta^n)$. Then, we need the number of Trotter steps to be $N_{\rm T} \sim N_{\rm terms} \norm{\widetilde{H}}^2 t^2 / \epsilon_{\rm Trotter} = \OO (N_\omega^n N_\beta^{n+1})$. Therefore the total number of gates is $N_{\rm G} = \OO (N_{\rm E} N_\omega^{2n} N_\beta^{2n+2})$. We note that a variety of methods have been devoloped to reduce the gate number in the Trotterisation algorithm~\cite{Berry2007, Wiebe2010, Berry2015}, which could be applied in our case.

In our algorithm, the system size can easily exceed the environment size. For example, to simulate the quantum master equation with $n=2$, considering an environment with a million discretised frequencies ($N_\omega = 10^6$) and a thousand interaction terms ($N_\beta = 10^3$), we only need about $30$ qubits for encoding the environment, which is even smaller than the system in a non-trivial quantum simulation problem (with above $50$ qubits).

\section{Thermalisation of a qubit on quantum computer}
\label{sec:thermalisation}

Let us consider the thermalisation of a qubit at the zero temperature and finite temperature. The system Hamiltonian is $\widetilde{H}_{\rm S} = -\frac{\Delta}{2} \sigma^{\rm z}$. The system is coupled to the environment via only one term, i.e.~$\widetilde H_{\rm I}=\alpha A \otimes \widetilde{B}$, where $A = \sigma^{\rm x}$, and $\widetilde{B}$ is the same as in Eq.~(\ref{eq:B}), but coupling coefficients are~\cite{Ritschel2014}
\begin{eqnarray}
g(\omega)=\left\{
\begin{array}{ll}
\sqrt{a\bar\gamma \frac{\abs{\omega}\bar\gamma}{\omega^2+{\bar\gamma}^2}}, & \omega \geq 0, \\
\sqrt{a\bar\gamma \frac{\abs{\omega}\bar\gamma e^{\beta\omega}}{\omega^2+{\bar\gamma}^2}}, & \omega < 0.
\end{array}
\right.
\end{eqnarray}
Here, $\beta = 1/k_{\rm B}T$ is the temperature, and $a$ and $\bar\gamma$ are constants with the dimensions of frequency. We take $\rho_{\rm S}(0) = \ketbra{+}{+}$ as the initial state of the qubit, and $\ket{+} = \frac{1}{\sqrt{2}}(\ket{0}+\ket{1})$.

In Fig.~\ref{fig:thermalisation}, we plot the probability in the ground state $p_{\rm g} = \bra{0}\rho_{\rm S}(t)\ket{0}$, where $\rho_{\rm S}(t)$ is the state of the qubit at time $t$. When the simulated environment is at the zero temperature, i.e.~$\beta \rightarrow +\infty$, the qubit evolves into the ground state $\ket{0}$, i.e.~$p_{\rm g}$ goes to $1$. When the temperature is finite, the probability approaches a finite value and coincides with the thermal distribution. For comparison, we also plotted the probability in the evolution driven by the corresponding Lindblad equation of the thermalisation~\cite{BreuerPetruccione}. The difference between the environment-simulation result and the Lindblad equation result is due to the discretisation of the environment spectrum and approximations used to derive the Lindblad equation, including neglecting high-order terms in TCL equation and the Markovian approximation.

\begin{figure}[tbph]
\includegraphics[width=1\linewidth]{\figpath 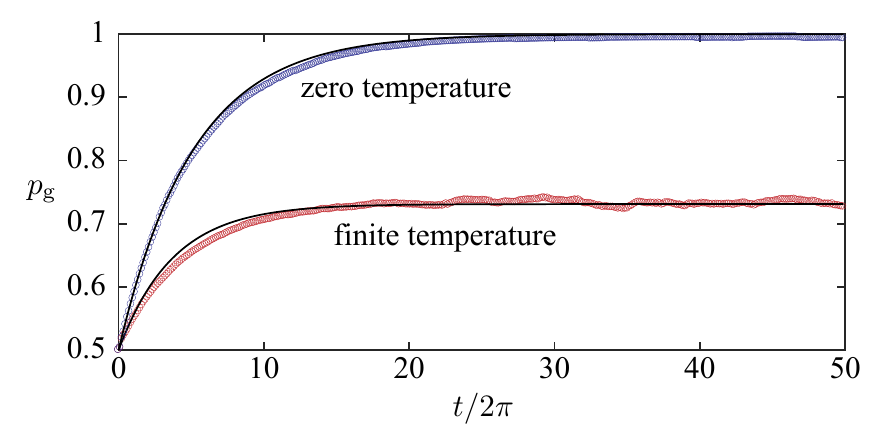}
\caption{
Probability in the ground state, $p_{\rm g}$. We take $\Delta = 1$, $\alpha\sqrt{a} = 0.01$, $\bar\gamma = 10$, $\delta\omega = 0.02$, $N_\omega = 401$. The dissipation of the environment is introduced using the condition reinitialisation protocol (see the end of Sec.~\ref{sec:CPDP}). The probabilities $p_{\rm g}$ for the zero temperature (blue circles) and the finite temperature ($\beta = 1$, red circles) are computed using the quantum trajectory approach~\cite{Gardiner2000, Carmichael1993}. We take $1000$ instances for the zero temperature and $5000$ instances for the finite temperature. The dissipation zone is $21 \leq x \leq 380$ for the zero temperature; further moving the dissipation zone towards $x = 0$ (e.g. taking the dissipation zone $11 \leq x \leq 390$) will change the correlation function. The dissipation zone is $3 \leq x \leq 398$ for the finite temperature, which is chosen to obtain the best fit to the Lindblad equation of the thermalisation. Black curves represent the result of corresponding Lindblad equation of the thermalisation.
}
\label{fig:thermalisation}
\end{figure}

\section{Conclusions}

In this paper, we propose a hardware-efficient quantum algorithm to simulate the TCL master equation up to any finite order. It is achieved by reproducing reservoir correlation functions using a minimised Hilbert space. The number of qubits representing the environment is $\sim \lfloor \frac{n}{2} \rfloor \log_2 (N_\omega N_\beta)$ in the $n$-th-order simulation. We remark that $n = 2$ in the simulation of the Markovian quantum master equation and the thermalisation. In our algorithm, the system size can easily exceed the environment size, e.g.~when the system has tens of qubits. Because the environment on the quantum computer is small, it needs to be reinitialised in the simulation of a long-time evolution. We also propose an efficient reinitialisation protocol without significantly changing reservoir correlation functions. We illustrate our algorithm by using a classical computer and numerically simulate the thermalisation of a qubit at the zero and finite temperatures. Our results pave the way for practical quantum open-system simulation on a universal quantum computer.

\begin{acknowledgments}
This work is supported by National Natural Science Foundation of China (Grant No. 11875050) and NSAF (Grant No. U1930403). HYS is also supported by China Postdoctoral Science Foundation (Grant No. 2018M630063) and National Natural Science Foundation of China (Grant No. 11905209).
\end{acknowledgments}

\appendix

\section{Gram-Schmidt orthogonalisation}

In this section, we explicitly present the Gram-Schmidt orthogonalisation process. We have $d_{n,{\rm max}}$ vectors $V_n = \{ \ket{\phi_{\Omega,m}(\cdots)} \}$ in $\mathcal{H}_{\rm E}^{\rm r}$, where $m \leq \lfloor n/2 \rfloor$. We label these vectors as $\ket{\phi_1}, \ldots, \ket{\phi_{d_{n,{\rm max}}}}$. Without loss of generality, we take $\ket{\phi_1} = \ket{\psi}$, which can simplify the preparation of the environment initial state on the quantum computer, and we assume that states from $\ket{\phi_1}$ to $\ket{\phi_{d_{\rm E}}}$ are linearly independent. We note that $g$ is a $d_{n,{\rm max}}$-dimensional matrix with rank $d_{\rm E}$. The state $\ket{\psi}$ is normalised, therefore we take $\ket{e_1} = \ket{\phi_1}$. Then, we can obtain $d_{\rm E}$ orthonormal basis states by iterating
\begin{eqnarray}
\ket{e_i} = \frac{\ket{\phi_i} - \sum_{j=1}^{i-1} \ket{e_j}\braket{e_j}{\phi_i}}{\norm{\ket{\phi_i} - \sum_{j=1}^{i-1} \ket{e_j}\braket{e_j}{\phi_i}}}.
\end{eqnarray}
Given $\ket{e_j} = \sum_{\phi \in V_n} e_{j,\phi} \ket{\phi}$, we compute the overlap using $\braket{e_j}{\phi_i} = \sum_{\phi \in V_n} e_{j,\phi}^* g_{\phi,\phi_i}$. The outcome of the Gram-Schmidt orthogonalization is the $d_{\rm E} \times d_{n,{\rm max}}$ matrix $e_{j,\phi}$.

Using the matrix $e_{j,\phi}$, we can express states $\ket{\phi} \in V_n$ and operators $b \in \{b_\beta(\omega)\}$ using the orthonormal basis of the subspace $\mathcal{H}_{\rm E}^{\rm r} = \mathrm{span}(V_n)$, i.e.~$\ket{\phi} = \sum_i \braket{e_i}{\phi} \ket{e_i}$ and $\Pi_{\rm E} b \Pi_{\rm E} = \sum_{i,j} \bra{e_i} b \ket{e_j} \ketbra{e_i}{e_j}$, where $\Pi_{\rm E} = \sum_{i=1}^{d_{\rm E}} \ketbra{e_i}{e_i}$ is the projection onto the subspace, $\braket{e_i}{\phi} = \sum_{\phi' \in V_n} e_{i,\phi'}^* g_{\phi',\phi}$ and $\bra{e_i} b \ket{e_j} = \sum_{\phi',\phi \in V_n} e_{i,\phi'}^* g_{\phi',\phi} e_{j,\phi}$.

Let $\{ \ket{\widetilde{e_i}} ~\vert~ i=1,\ldots,d_{\rm E} \}$ be $d_{\rm E}$-dimensional orthonormal states. Each $\ket{\widetilde{e_i}}$ is a state in $\widetilde{\mathcal{H}}_{\rm E}$ on the quantum computer. Then, for operators $b \in \{b_\beta(\omega)\}$, we define $\widetilde{b} \equiv \sum_{i,j} \bra{e_i} b \ket{e_j} \ketbra{\widetilde{e}_i}{\widetilde{e}_j}$. Because $b_\beta^\dag(\omega) = b_\beta(-\omega)$, $[\Pi_{\rm E} b_\beta(\omega) \Pi_{\rm E}]^\dag = \Pi_{\rm E} b_\beta(-\omega) \Pi_{\rm E}$. Therefore, $\widetilde{b}_\beta^\dag(\omega) = \widetilde{b}_\beta(-\omega)$.

For a state $\ket{\varphi} \in \mathrm{span}(V_n)$, we define $\ket{\widetilde{\varphi}} \equiv \sum_i \braket{e_i}{\varphi} \ket{\widetilde{e}_i}$. Then,
\begin{eqnarray}
\bra{\widetilde{\varphi}} \widetilde{b}_m \cdots \widetilde{b}_2 \widetilde{b}_1 \ket{\widetilde{\varphi}} = \bra{\varphi} b_m \Pi_{\rm E} \cdots \Pi_{\rm E} b_2 \Pi_{\rm E} b_1 \ket{\varphi}.
\end{eqnarray}
We remark that $\Pi_{\rm E} \ket{\varphi} = \ket{\varphi}$.

Because $\Pi_{\rm E} b_m \Pi_{\rm E} \cdots \Pi_{\rm E} b_2 \Pi_{\rm E} b_1 \ket{\psi} = b_m \cdots b_2 b_1 \ket{\psi} \in V_n$ for all $m \leq \lfloor n/2 \rfloor$, the following equation holds for all $m\leq 2\lfloor n/2 \rfloor +1$,
\begin{eqnarray}
\bra{\psi} b_m \Pi_{\rm E} \cdots \Pi_{\rm E} b_2 \Pi_{\rm E} b_1 \ket{\psi} = \bra{\psi} b_m \cdots b_2 b_1 \ket{\psi}.
\end{eqnarray}
Therefore, Eq.~(\ref{eq:vbbbv}) holds for all $m\leq 2\lfloor n/2 \rfloor +1$.

\section{Proof of the algorithm}

Because $\bra{\widetilde{\phi}_{\Omega_{\rm L},m_{\rm L}}(\cdots)}\widetilde{b}_\beta(\omega)\ket{\widetilde{\phi}_{\Omega_{\rm R},m_{\rm R}}(\cdots)} = 0$ if $\omega \neq \Omega_{\rm L} - \Omega_{\rm R}$, we have $\widetilde{b}_\beta(\omega) = \sum_{\Omega_{\rm L} - \Omega_{\rm R} = \omega} \widetilde{\Pi}_{\Omega_{\rm L}} \widetilde{b}_\beta(\omega) \widetilde{\Pi}_{\Omega_{\rm R}}$, then $\widetilde{B}_\beta(t) = \sum_\omega e^{-i\omega t} \widetilde{b}_\beta(\omega)$. Therefore, $\widetilde{B}_\beta(\omega) = \widetilde{b}_\beta(\omega)$.

Correlation functions on the quantum computer can be expressed as
\begin{eqnarray}
&& \Tr \big[ \widetilde{\BB}_{\beta}(\omega,\nu) \cdots \widetilde{\BB}_{\beta_{m-1}}(\omega_{m-1},\nu_{m-1}) \widetilde{\rho}_{\rm E} \big] \notag \\
&=& \bra{\widetilde{\psi}} [\widetilde{b}_{\beta_{m-1}}(\omega_{m-1})]^{1-\nu_{m-1}} \cdots [\widetilde{b}_{\beta}(\omega)]^{1-\nu}
\notag \\
&&\times [\widetilde{b}_{\beta}(\omega)]^{\nu} \cdots [\widetilde{b}_{\beta_{m-1}}(\omega_{m-1})]^{\nu_{m-1}} \ket{\widetilde{\psi}}.
\end{eqnarray}
Because of Eq.~(\ref{eq:vbbbv}), the following equation holds for all $m \leq n$,
\begin{eqnarray}
&& \Tr \big[ \widetilde{\BB}_{\beta}(\omega,\nu) \cdots \widetilde{\BB}_{\beta_{m-1}}(\omega_{m-1},\nu_{m-1}) \widetilde{\rho}_{\rm E} \big] \notag \\
&=& \bra{\psi} [b_{\beta_{m-1}}(\omega_{m-1})]^{1-\nu_{m-1}} \cdots [b_{\beta}(\omega)]^{1-\nu}
\notag \\
&&\times [b_{\beta}(\omega)]^{\nu} \cdots [b_{\beta_{m-1}}(\omega_{m-1})]^{\nu_{m-1}} \ket{\psi} \notag \\
&=& \Tr \big[ \BB_{\beta}(\omega,\nu) \cdots \BB_{\beta_{m-1}}(\omega_{m-1},\nu_{m-1}) \rho_{\rm E} \big].
\end{eqnarray}
Therefore, Eq.~(\ref{eq:BBrho}) holds for all $m \leq n$.

\section{Norm of $\widetilde{B}_{\beta}$}

For any state $\ket{\widetilde{\varphi}} \in \widetilde{\mathcal{H}}_{\rm E}$, we have
\begin{eqnarray}
&& \norm{\widetilde{B}_{\beta} \ket{\widetilde{\varphi}}}^2 = \bra{\varphi} b_{\beta}^\dag \Pi_{\rm E} b_{\beta} \ket{\varphi} \notag \\
&\leq & \norm{b_{\beta} \ket{\varphi}}^2 \leq \norm{b_{\beta}}^2 \norm{\ket{\varphi}}^2,
\end{eqnarray}
where $b_\beta = \sum_\omega b_{\beta}(\omega) = B_\beta\otimes\openone_{\rm a}$. Notice that $\norm{\ket{\varphi}} = \norm{\ket{\widetilde{\varphi}}}$ and $\norm{b_{\beta}} = \norm{B_{\beta}}$, we have $\norm{\widetilde{B}_{\beta}} \leq \norm{B_{\beta}}$.

\section{Correlation functions reproduced in the fourth-order example}

Correlation functions reproduced in the environment given by Eq.~(\ref{eq:fourth-order_H}) and Eq.~(\ref{eq:fourth-order_B}) are
\begin{eqnarray}
\bra{\rm v} \widetilde{B}_\beta(s) \widetilde{B}_{\beta_1}(s_1) \ket{\rm v} = \sum_\omega e^{-i\omega(s-s_1)} g_{\beta,\omega} g_{\beta_1,\omega}^*,
\end{eqnarray}
\begin{eqnarray}
\bra{\rm v} \widetilde{B}_\beta(s) \widetilde{B}_{\beta_1}(s_1) \widetilde{B}_{\beta_2}(s_2) \ket{\rm v} = 0
\end{eqnarray}
and
\begin{eqnarray}
&& \bra{\rm v} \widetilde{B}_\beta(s) \widetilde{B}_{\beta_1}(s_1) \widetilde{B}_{\beta_2}(s_2) \widetilde{B}_{\beta_3}(s_3) \ket{\rm v} \notag \\
&=& \bra{\rm v} \widetilde{B}_\beta(s) \widetilde{B}_{\beta_1}(s_1) \ket{\rm v} \bra{\rm v} \widetilde{B}_{\beta_2}(s_2) \widetilde{B}_{\beta_3}(s_3) \ket{\rm v} \notag \\
&&+ \sum_{\omega,\omega',\omega_1,\omega_2} e^{-i\omega'(s-s_1)} e^{-i(\omega_1+\omega_2)(s_1-s_2)} e^{-i\omega(s_2-s_3)} \notag \\
&&\times g_{\beta,\omega'} g_{\beta_1,\omega',\omega_1,\omega_2} g_{\beta_2,\omega,\omega_1,\omega_2}^* g_{\beta_3,\omega}^*.
\end{eqnarray}

\end{document}